\documentclass[aps,floats,preprintnumbers,nofootinbib]{revtex4}
\usepackage{amsmath,amssymb}
\usepackage{epsfig}

\def\rts{\sqrt s}
\def\lsim{\mathrel{\raise.3ex\hbox{$<$\kern-.75em\lower1ex\hbox{$\sim$}}}}
\def\gsim{\mathrel{\raise.3ex\hbox{$>$\kern-.75em\lower1ex\hbox{$\sim$}}}}

\newcommand{\nc}{\newcommand}
\nc{\baa}{\begin{array}}     \nc{\eaa}{\end{array}}
\nc{\bew}{\beta_W}
\nc{\dtz}{d^2_{0,0}}
\nc{\dzz}{d^0_{0,0}}
\nc{\dto}{d^2_{1,0}}

\def\mhmax{\mh^{\rm max}}
\def\tev{~{\rm TeV}}
\def\gev{~{\rm GeV}}

\def\mplanck{M_{\rm Pl}}
\def\lcut{\overline \Lambda}
\def\bea{\begin{eqnarray}}
\def\eea{\end{eqnarray}}
\def\non{\nonumber}
\def\lsp{\;\;\;\;\;}
\def\beq{\begin{equation}}
\def\eeq{\end{equation}}
\def\eg{{\it e.g.}}

\def\re{{\rm Re}~}
\def\ocal{{\cal O}}
\def\ct{\cos\theta}
\def\cts{\cos^2\theta}
\def\lamw{\lwh}
\def\wlwl{\wl^+\wl^-\to \wl^+\wl^-}
\def\mws{\mw^2}
\def\lamws{\lwh^2}
\def\mgs{m_G^2}
\def\cws{\cos^2\theta_W}
\def\ms{\overline{m}_{\rm scal}^2}
\def\bce{\begin{center}}
\def\ece{\end{center}}

\def\br{{\rm BR}}
\def\epem{e^+e^-}
\def\mupmum{\mu^+\mu^-}
\def\mplv{M_{Pl\,5}}

\def\phio{\phi_0}

\def\anti{\overline}
\def\mz{m_Z}
\def\mw{m_W}

\def\gam{\gamma}

\def\mh{m_{h}}
\def\mphi{m_\phi}
\def\what{\widehat}
\def\lwh{\widehat\Lambda_W}

\def\lphi{\Lambda_\phi}
\def\lbar{\anti\Lambda}

\def\mphi{m_\phi}

\def\hbar{\overline h}

\def\mpl{M_{Pl}}
\def\mplanck{\mpl}
\def\wl{W_L}
\def\awl{{\cal A}_{\wlwl}}
\def\lnda{\Lambda_{\rm NDA}}
\def\call{{\cal L}}

\def\mpl{m_{Pl}}
\def\mompl{m_0/\mpl}
\preprint{UCD-2006-15}
\preprint{IFT-06-19}

\begin{document}
\title{KK Gravitons and Unitarity Violation in the Randall-Sundrum Model}
\author{Bohdan Grzadkowski}
\affiliation{Institute of Theoretical Physics  \\
   Warsaw University, 00-681 Warsaw, Poland}

\author{John F. Gunion}
\affiliation{ Davis Institute for High Energy Physics \\
University of California, Davis, CA 95616, U.S.A.}

\begin{abstract}
  We show that perturbative unitarity for $\wlwl$ scattering places
  significant constraints on the Randall-Sundrum 
  theory with two 3-branes, with matter confined to the TeV brane.
  The exchange of massive 4D Kaluza-Klein gravitons leads to
  amplitudes growing linearly with the CM energy squared. Summing over
  KK gravitons up to a scale $\lbar$ and testing unitarity at
  $\rts=\lbar$, one finds that unitarity is violated for
  $\lbar$ below the 'naive
  dimensional analysis' scale, $\lnda$. We evaluate $\lbar$ as
  a function of the curvature ratio $\mompl$ for the pure gravity
  theory. We then demonstrate that unitarity need not be violated at
  $\lbar$ in the presence of a heavy Higgs boson.
  In fact, much larger Higgs masses are consistent with unitarity than if no KK
  gravitons are present. Observation of the mass and width (or cross
  section) of one or more KK gravitons at the LHC will directly
  determine $\mompl$ and the scale $\lwh$ specifying the couplings of
  matter to the KK gravitons. With this information in hand and a
  measurement of the Higgs boson mass, one can
  determine the precise scale $\lbar$ below which unitarity will
  remain valid.

\end{abstract}

\maketitle

\vspace*{-.3in}
\section{Introduction}
\label{intro}
\vspace*{-.1in}

The Standard Model (SM) of electroweak interactions is confirmed 
by all existing experimental data. However the model
suffers from certain theoretical drawbacks. One of these is the hierarchy
problem: namely, the SM can not consistently accommodate the weak
energy scale ${\cal O}(1\tev)$ and a much higher scale such as the
Planck mass scale ${\cal O}(10^{19}\gev)$.  Therefore, it is commonly
believed that the SM is only an effective theory emerging as the
low-energy limit of some more fundamental high-scale theory that
presumably could contain gravitational interactions. 

Models that involve extra spatial dimensions could provide a solution
to the hierarchy problem. One attractive proposal was formulated by
Randall and Sundrum~(RS) ~\cite{Randall:1999ee}. They postulate a 5D
universe with two 4D surfaces (``3-branes''). All the SM particles and
forces with the exception of gravity are assumed to be confined to one
of those 3-branes called the visible or TeV brane.  Gravity lives on
the visible brane, on the second brane (the ``hidden brane'') and in
the bulk.  All mass scales in the 5D theory are of order of the Planck
mass.  By placing the SM fields on the visible brane, all the order
Planck mass terms are rescaled by an exponential suppression factor
(the ``warp factor'') $\Omega_0\equiv e^{-m_0 b_0/2}$, which reduces
them down to the weak scale ${\cal O}(1 \tev)$ on the visible brane
without any severe fine tuning. To achieve the necessary suppression,
one needs $m_0 b_0 /2 \sim 35$. This is a great improvement compared
to the original problem of accommodating both the weak and the Planck
scale within a single theory. 
The RS model is defined by the 5-D action:
\bea
S&=&-\int d^4x\, dy \sqrt{-\what g}\left({2\mplv^3\what R }+\Lambda\right)\non\\
&&+\int d^4x\,\sqrt{-g_{\rm hid}}({\cal L}_{\rm hid}-V_{\rm hid})
+\int d^4x\,\sqrt{-g_{\rm vis}}({\cal L}_{\rm vis}-V_{\rm vis})\,,
\eea
where the notation is self-explanatory, see also \cite{Dominici:2002jv} for details.
In order to obtain a consistent solution to Einstein's
equations corresponding to a low-energy effective theory that is flat,
the branes must have
equal but opposite cosmological constants and these must
be precisely related to the bulk cosmological constant;
$V_{\rm hid}=-V_{\rm vis}={24\mplv^3 m_0 }\lsp$ and
$\lsp \Lambda=-{24\mplv^3 m_0^2}$.
With these choices, the following metric is a solution of Einstein's equations:
\beq
\what g_{\what\mu\what\nu}(x,y)=\left(
\begin{array}{ccc} 
e^{-2m_0 b_0 |y|}\eta_{\mu\nu}&|& 0 \\
\hline 
0 &|& -b_0^2
\end{array}
\right) \,.
\eeq
After an expansion around the background metric we obtain the gravity-matter interactions
\beq
\call_{\rm int}=-{1\over\lwh}\sum_{n\neq 0} h_{\mu\nu}^n T^{\mu\nu} -{\phi_0\over\lphi}T_\mu^\mu
\label{int}
\eeq
where $h_{\mu\nu}^n(x)$ are the Kaluza-Klein (KK) modes (with mass
$m_n$) of the graviton field $h_{\mu\nu}(x,y)$, $\phio(x)$ is the
radion field (the quantum degree of freedom associated with
fluctuations of the distance between the branes), $\lwh \simeq \sqrt 2
\mpl \Omega_0$, where $\Omega_0 = e^{-m_0 b_0/2}$, and $\lphi = \sqrt
3\,\lwh$.  To solve the hierarchy problem, $\lwh$ should be of order
$1-10\tev$, or perhaps
higher~\cite{Randall:1999ee}. Note from Eq.~(\ref{int}) that the
radion couples to matter with coupling strength $1/\lphi$.  In
addition to the radion, the model contains a conventional Higgs boson,
$h_0$.  The RS model solves the hierarchy problem by virtue of the
fact that the 4D electro-weak scale is given in terms of the ${\cal
  O}(\mpl)$ 5D Higgs vev, $\what v$, by: \beq v_0= \Omega_0 \what v=
e^{-m_0 b_0/2} \what v \sim 1\tev \lsp {\rm for}\lsp m_0 b_0 /2 \sim
35\,.  \eeq

However, the RS model is trustworthy in its own right 
only if the 5D curvature $m_0$ is small compared to the 5D Planck
mass, $\mplv$~\cite{Randall:1999ee}.  The reason is that for higher
$m_0$ one can't trust the RS solution of Einstein's equations since
then $m_0$, a parameter of the solution, is greater than the scale up
to which classical gravity can be trusted.  The $m_0<\mplv$
requirement and the fundamental RS relation $\mpl^2=2\mplv^3/m_0$
imply that $\mompl=2^{-1/2}(m_0/\mplv)^{3/2}$ should be significantly
smaller than 1.  Roughly, it is believed that $\mompl\lsim 0.1$ is
required for internal consistency of the RS 5D model.  String theory
estimates are usually smaller, typically of order $\mompl\sim
0.01$~\cite{Davoudiasl:1999jd}.  At the same time, the effective 4D RS
theory should be well behaved up to some maximum energy that can be
estimated in a number of ways.  One estimate of the maximum energy
scale is that obtained using the 'naive dimensional analysis' (NDA)
approach~\cite{Manohar:1983md}, and the associated scale is denoted
$\lnda$.~\footnote{The 4D condition for the cutoff $\lnda$ (which
  corresponds to the scale at which the theory becomes strongly
  coupled) is $(\lnda/\lwh)^2 N/(4\pi)^2\sim 1$, where $N$ is the
  number of KK-gravitons lighter than $\lnda$ (implying that they
  should be included in the low-energy effective theory). For the RS
  model the graviton mass spectrum for large $n$ is $m_n\simeq m_0\pi
  n \Omega_0$, implying $N\sim \lnda/(m_0\pi\Omega_0)$ which leads to
  Eq.~(\ref{nda}).}  One finds
\beq
\lnda= 2^{7/6}\pi (\mompl)^{1/3}\lwh\,,
\label{nda}
\eeq
where $\lwh$ was defined in Eq.~(\ref{int}); its inverse sets the
strength of the coupling between matter and gravitons.  We emphasize
that $\lnda$ is obtained when the exchange of the whole tower of KK
modes up to $\lnda$ is taken into account.~\footnote{An equivalent
  expression for $\lnda$ is $\lnda = 2^{5/3} \pi \Omega_0 \mplv$, as
  is consistent with redshifting the 5d cutoff $\mplv$ to the TeV
  brane.}  Physically, $\lnda$ is the energy scale at which the theory
starts to become strongly coupled and string/$M$-theoretic excitations
appear from a 4D observer's point of view ~\cite{Randall:1999ee}.
Above $\lnda$, the RS effective theory is expected to start to break
down and additional new physics must emerge.  An interesting question
is whether the model violates other theoretical constraints at this
same scale, a lower scale, or if everything is completely consistent
for energy scales below $\lnda$.  In this paper, we show that
unitarity in the $J=0$ partial wave of $\wlwl$ scattering is always
violated in the RS model for energies below the $\lnda$ scale.  We
will define $\lbar$ as the largest $\rts$ value such that if we sum
over graviton resonances with mass below $\lbar$ (but do not include
diagrams containing the Higgs boson or radion of the model) then
$\wlwl$ scattering remains unitary in the $J=0,1,2$ partial waves. We
find that $\lbar$ depends upon the curvature ratio, $\mompl$, in much
the same manner as $\lnda$, but is always $<\lnda$. The latter implies
that $\lbar$ is a more precise estimate of the energy scale at which
the theory becomes strongly interacting.

The maximum energy scale determined from $\wlwl$ scattering unitarity
if a light Higgs boson is included is essentially the same as $\lbar$
(and is essentially independent of the radion mass assuming no Higgs
radion mixing). However, the presence of a heavy Higgs boson can delay
the onset of unitarity violation in $\wlwl$ to energies above $\lbar$.
As a result, the question arises as to whether we should continue to
cutoff our KK sums at a maximum mass equal to the $\lbar$ obtained for
the $J=0$ partial wave before inclusion of Higgs and radion
contributions. Alternative choices include using the $\lbar$ obtained
using the $J=1$ or $J=2$ partial waves, for which Higgs and radion
exchanges are less important; see later plots in Fig.~\ref{lndaratio}.
One could also consider using the $\lbar$ determined by unitarity
violation of the scattering amplitudes for transversely polarized
vector bosons, where it is known~\cite{Han:2004wt} that the leading
contribution ($\propto s$) originates purely from graviton 
exchange.~\footnote{
  In other words, for graviton exchange,
  in contrast to gauge theories, amplitudes grow as $s$ both for
  longitudinal and transverse polarization of the vector bosons
  involved. For fermions in either the initial or final state,  
  amplitudes rise as $s^{1/2}$.} 
Each of these choices yields a somewhat different maximum $\rts$ value
(always modestly higher than the $\lbar$ obtained for $J=0$ and KK
exchanges only). We have opted to always cut off our sums over KK
resonances for masses above the $\lbar$ obtained for $J=0$ with KK
exchanges only. This will provide a conservative estimate of the
influence of KK resonance exchanges on unitarity limits.

As will be discussed later, observation of
even one KK graviton and its width or hadron-collider cross section
will determine both $\lwh$ and $\mompl$, from which $\lbar$ can be determined. 
If the Higgs boson mass is
also known, one can then make a fairly precise
determination of the $\rts$ scale for which unitarity constraints are
still obeyed for the given $\lwh$, $\mompl$ and Higgs mass
and how this scale relates to $\lbar$. 

In this paper, we will not consider the possible extension of the
RS model obtained by including mixing between
gravitational and electroweak degrees of freedom~\cite{Giudice:2000av,
Dominici:2002jv}. These can substantially amplify the radion
contribution to 
$WW$ scattering (see \cite{Grzadkowski:2005tx}). However
in the absence of such mixing, the $\phi_0$ and $h_0$ are
mass eigenstates,~\footnote{We assume that there is a mechanism that
stabilizes the inter-brane distance providing a mass for the radion. The simplest
scenario is the one with a bulk scalar field~\cite{Goldberger:1999wh} (see also 
\cite{Grzadkowski:2003fx}) which is an 
$SU(2)$ singlet and therefore does not influence $WW$ scattering.}
which we denote as $\phi$ and $h$.  An important
parameter is the quantity
\beq
R^2\equiv \what g_{VVh}^2+\what g_{VV\phi}^2= \what g_{ffh}^2+\what g_{ff\phi}^2=1+\gam^2\,,
\eeq
where the $\what g~$'s are defined~\cite{Dominici:2002jv} relative to SM Higgs coupling
strength (\eg\ $\what g_{WWh}=g_{WWh}/(g\mw)$) and $\gam\equiv v_0/\lphi$ is $\ll 1$ 
for typical $\lphi$ choices (with $v_0=246\gev$).

 The $h$, the $\phi$ and the KK gravitons (generically denoted $G$)
must all be considered in computing the high energy behavior of
a process such as $\wlwl$ scattering. 
As usual, there is a cancellation between  scalar ($h$ and $\phi$) 
exchanges and gauge boson
exchanges that leads to an amplitude $\awl$ that obeys~\footnote{The presence of the
radion spoils the cancellation of terms $\propto s$. However in the absence of
 radion-Higgs  
mixing those effects are numerically irrelevant~\cite{Grzadkowski:2005tx};
see also Eq.~(\ref{ao}) and below.} unitarity 
constraints (in particular, $|{\rm Re}a_0|\leq 1/2$ for the $J=0$
partial wave) so long as $m_h\lsim 870\gev$.
However, each
KK resonance with mass below $\rts$ will give a contribution to $\awl$ that grows with
$s$, and, since the number of such KK resonances increases
as the energy increases, their net effect does not decouple,
and, in fact, becomes increasingly important as $\rts$ increases. 
Thus, unitarity can easily
be violated for rather modest energies. Unitarity in the context
of the RS model has also been discussed in \cite{Han:2001xs} and
\cite{Choudhury:2001ke}.

The paper is organized as follows.
Sec.~\ref{ww} presents leading analytical results for
the partial wave amplitudes in the context of the RS model.
Sec.~\ref{params} 
discusses the parameters of the model, including graviton widths,
and experimental limits on these parameters.
Sec.~\ref{numerics} is devoted to detailed numerical analysis.
Sec.~\ref{experiment} discusses the means for
using future experiments to determine the model parameters.
A summary and some concluding remarks  are given in Sec.~\ref{sum}. 

\vspace*{-.2in}
\section{Vector boson scattering and KK exchanges}
\label{ww}
\vspace*{-.1in}

Let us begin by reviewing the 
limit on the Higgs-boson mass in the SM obtained by requiring that  
$\wlwl$ scattering be unitary
at high energy. The constraint arises when we consider
the elastic scattering of longitudinally polarized $W$ bosons.
The amplitude can be decomposed into partial wave contributions:
$T(s, \cos\theta) = 16 \pi \sum_J (2J+1)a_J(s) P_J(\cos\theta)$, where
$a_J(s) = \frac{1}{32 \pi(2J+1)} \int_{-1}^{1}T(s,\cos\theta)
P_J(\cos\theta) d\cos\theta\,.
$
In the SM, the partial wave amplitudes take the asymptotic form
$a_J = A_J \left(\frac{s}{m_W^2}\right)^2 + B_J \left(\frac{s}{m_W^2}\right) + C_J\,,
$
where $s$ is the center-of-mass energy squared.
Contributions that are divergent in the limit $s\to\infty$ appear only
for $J=0$, 1 and 2. 
The $A$-terms vanish by virtue of gauge invariance, while, as is very well known,
the $B$-term for $J=1$ and $0$~($B_2=0$) 
arising from gauge interaction diagrams is canceled by Higgs-boson exchange
diagrams. In the high-energy limit, the result is that
$a_J$ asymptotes to an  $\mh $-dependent constant. Imposing the
unitarity limit of $|\re a_J|<1/2$  
implies the Lee-Quigg-Thacker bound~\cite{Lee:1977eg} 
for the Higgs boson mass: $\mh  \lsim 870 \gev$.

We will show that within the RS model $\wlwl$ scattering violates
unitarity in the $J=0$ partial wave at an energy scale below $\lnda$ 
when KK graviton exchanges are included.  The various contributions to
the amplitude are given in Table~\ref{ampww}.  From the table, we see
that
in the SM, obtained by setting $R^2=1$, the gauge boson contributions
and Higgs exchange contributions cancel at $\ocal(s^2)$ and
$\ocal(s^1)$. Regarding $G$ exchange contributions, we note that the
apparent singularity in the $\cos\theta$ integral of the leading
$\ocal(s)$ t-channel $G$ exchange is regularized by the graviton mass
and width (neglected in the table).

\begin{table}[t!]
\begin{tabular}{|c||c|c|}
\hline
diagram & $\ocal(\frac{s^2}{v^4})$ & $\ocal(\frac{s^1}{v^2})$ \\
\hline\hline
$\gamma, Z$ s-channel & $-\frac{s^2}{g^2v^4} 4 \ct$ & $-\frac{s}{v^2} \ct$  \\
\hline
$\gamma, Z$ t-channel & $-\frac{s^2}{g^2v^4}(-3+2\ct+\cts)$ & $-\frac{s}{v^2}\frac32 (1-5\ct)$ \\
\hline
$WWWW$ contact   & $-\frac{s^2}{g^2v^4}(3-6\ct-\cts)$ & $-\frac{s}{v^2} 2(-1+3\ct)$ \\
\hline 
$G$ s-channel         & $0$ & $-\frac{s}{24\lamw^2}(-1+3\cts)$ \\
\hline
$G$ t-channel         & $0$ & $-\frac{s}{24\lamw^2}\frac{13+10\ct+\cts}{-1+\ct}$ \\
\hline
$(h-\phi)$ s-channel    & $0$ & $-\frac{s}{v^2}R^2$ \\
\hline
$(h-\phi)$ t-channel    & $0$ & $-\frac{s}{v^2}\frac{-1+\ct}{2} R^2$ \\
\hline
\end{tabular}
\vspace*{-.07in}
\caption{The leading contributions to the $\wlwl$ 
amplitude, where $R^2$ is defined in the text.
$G$ denotes a single KK graviton.}
\label{ampww}
\vspace*{-.18in}
\end{table}

It is worth noting that even though the graviton exchange amplitude
has the same amplitude growth $\propto s$ as the SM vector boson and
contact interactions, its angular dependence is different ($J=2$ vs.
$J=1$); therefore, the graviton cannot act in place of the Higgs boson
to restore correct high-energy unitary behavior for $a_0$ and $a_1$.
It is also noteworthy that the $t$-channel graviton contributions to the
$a_J$ are quite substantial as a result of the $(1-\cos\theta)^{-1}$
structure that is regulated by the graviton mass and width.

As is well known, the cancellation of the $\ocal(s^2)$ contributions
in Table~\ref{ampww} between the contact term and $s$- and $t$-channel
gauge-boson exchange diagrams is guaranteed by gauge invariance. Even
more remarkable is the cancellation of the most divergent graviton
exchange terms.  Indeed, a naive power counting shows that the
graviton exchange can yield terms at $\ocal(s^5)$, while the actual
calculation shows that only the linear term $\propto s$ survives; all
the terms with faster growth of $\ocal(s^5,s^4,s^3,s^2)$ cancel.  The
mechanism behind the cancellation is as follows.  In the high-energy
region the massive graviton propagator grows with energy as $k_\mu
k_\nu k_\alpha k_\beta k^{-2}$, where $k$ is the momentum carried by
the graviton, which will be of order $\rts$.  The graviton couples to the
energy-momentum tensor $T_{\mu\nu}$, so the amplitude for a single
graviton exchange is of the form
$T_{\mu\nu}D^{\mu \nu,\alpha \beta}T_{\alpha\beta}\,.$
Since the energy-momentum tensor is conserved, $k^\mu T_{\mu\nu}=0$,
terms in the numerator of the graviton propagator
proportional to the momentum don't contribute.  (Note that
for this argument to apply, all the external particles must be on
their mass shell.) This removes two potential powers of $s$ in the
amplitude.  In order to understand the disappearance of two additional
powers of $s$, let us calculate the energy-momentum tensor for the
final state consisting of a pair of longitudinal $W$ bosons.  A direct
calculation reveals the following form of the $4\times 4$ tensor:
\beq
\langle 0|T^{\mu\nu}|\wl^+\wl^-\rangle =
\left(
\baa{cccc}
0&0&0&0\\
0&\frac16[(1-2\bew)\dzz + 2(\bew-2)\dtz]s&0&-\frac{1}{\sqrt{6}} (s+4\mws)\dto\\
0&0&-\frac12 s \dzz&0\\
0&-\frac{1}{\sqrt{6}} (s+4\mws)\dto&0&-\frac16[(1+\bew)\dzz + 2(\bew-2)\dtz]s \,,
\eaa
\right)
\label{tmunu}
\eeq
in the reference frame in which the off-shell graviton is at rest.
The scattering angle is measured relative to the direction of motion
of the $W^-$, $d^J_{\mu\mu'}(\ct)=d^J_{\mu\mu'}$ stands for the Wigner
$d$ function and $\bew\equiv 1- 4\mws/s$.  Note that the factor
$1/\mws$, which comes from the vector boson polarization vectors, has
been canceled by two powers of $\mw$ coming from on-shellness of the
longitudinal vector bosons, i.e. $\mws$ replaces an $s$ that
originates from $T^{\mu\nu}$. In short, when the two
vertices are contracted with the propagator of the virtual graviton,
four potential powers of $s$ disappear leading to a single power of
$s$. These arguments apply equally to
$s$- and $t$-channel diagrams.

From the terms $\propto s$ and $\propto constant$ in the amplitude,
one obtains the leading terms in the partial wave amplitudes that are
$\propto s$, $\propto \ln s$, and $\propto constant$.  We give below
the leading terms deriving from a single KK graviton, the SM vector
bosons and the $\phi-h$ exchanges (for the $J=2$, $1$ and $0$ partial
waves):
\vspace*{-.1cm}
\bea
 a_2 & = & -\frac{1}{960\pi\lamws}\Biggl\{\left[ 91 + 30\log \left(\frac{\mgs}{s}\right) 
\right] s +
\left[241 + 210\log\left(\frac{\mgs}{s}\right)\right] \mgs 
+ 32 g^2v^2\Biggr\} + \ocal(s^{-1}) \label{wwt} \\
a_1 & = & -\frac{1}{1152\pi\lamws}\Biggl\{\left[73 + 
36\log\left(\frac{\mgs}{s}\right)\right]s +
36 \left[1 + 3\log\left(\frac{\mgs}{s}\right)\right]\mgs + 37
g^2v^2\Biggr\}
\non\\ && 
\quad\qquad+\frac{1}{96\pi}\left[\frac{s}{v^2}(1-R^2)
-R^2g^2 + 
\frac{12\cws-1}{2\cws}g^2\right] + \ocal(s^{-1})\non\\
\label{wwo} \\ [-0.3cm]
a_0 & = &  -\frac{1}{384 \pi\lamws}\Biggl\{\left[11 + 12 
\log\left(\frac{\mgs}{s}\right)\right]s
-\left[10 - 12\log\left(\frac{\mgs}{s}\right)\right]\mgs + 19 g^2 v^2 \Biggr\}  \non \\ 
 &&\qquad\quad + \frac{1}{32\pi}\left[\frac{s}{v^2}(1-R^2)
 + R^2 g^2
- 4\frac{\ms}{v^2}\right] + \ocal(s^{-1})\,,
\label{wwz}
\eea
where $\ms=\what g_{VVh}^2\mh^2+\what g_{VV\phi}^2\mphi^2$ is equal to
$\mh^2+\gam^2\mphi^2$ in the absence of Higgs-radion mixing.
It is amusing to note that in the case of $a_2$, the $s$-channel
contribution is quite minor, contributing just $a_2\ni
-\frac{1}{960\pi\lamws}$ to Eq.~(\ref{wwt}).
Of course, one should sum over all relevant KK gravitons.
We include all KK states with $m_n\leq \lcut$, where, as stated earlier,
$\lcut$ will be taken to be the largest energy or mass scale for which
$\wlwl$ scattering remains unitarity in all partial waves before
taking into account $h$ and $\phi$ exchanges.  It
is important to 
note that in our calculations the full sum over all modes with
$m_n\leq \lcut$ will be 
included even when considering $\rts$ values above or below $\lcut$ .

In our numerical
results, we employ exact expressions for the $h$, $\phi$ and all
KK contributions to $a_{0,1,2}$. Nonetheless, some analytic
understanding is useful.  We focus on $a_0$.
Eq.~(\ref{wwz}) shows that the Higgs plus gauge boson
contributions in the SM limit (obtained by taking $R^2=1$ and $\ms=\mh^2$) combine to give a negative constant value at large
$s$. If we add the radion, but neglect the KK graviton exchanges, then
the leading terms for $a_0$ are:
\beq
a_0=
\frac{1}{32\pi}\left[f(s) + g^2R^2 - 4 \frac{\ms}{v^2} \right],\,
\label{ao}
\eeq
where~\cite{Dominici:2002jv,Han:2001xs} $f(s)=\frac{s}{v^2}(1-R^2) =
-\frac{s}{\lphi^2}$ in the absence of Higgs-radion mixing.  
The negative signs for $f(s)$ and in front of $\ms$ imply some amplification
of unitarity-violation in the $\sim\tev$ energy range as compared to
that which is present in the SM for large $\mh$ values.
However, given the $1/(32\pi)$ factor and the fact that  we typically consider $\rts$ values of order
$\lnda,\lwh,\lphi$ or below, this
residual unitarity-violating behavior is never a dominant effect
when the Higgs-radion mixing~\cite{Grzadkowski:2005tx} is neglected.
Ultimately, at higher $\rts$ values near $\lcut$
it is usually the purely KK graviton exchanges that dominate.
From the asymptotic formula for $a_0$, a KK graviton
with mass significantly below $\rts$  enters $\re a_0$ 
with a positive sign.  The sum of all these
contributions is very substantial if $\mompl$ is small.

\vspace*{-.1in}
\section{Parameters of the Model and Experimental Limits}
\label{params}
\vspace*{-.1in}

At this point, it is useful to specify more fully the characteristics
of the KK graviton excitations. We start with the
parameters of the RS model: $\lwh \simeq \sqrt 2 \mpl \Omega_0$
(and $\lphi=\sqrt 3\lwh$) and the curvature $m_0$.
In terms of these, one finds $m_n=m_0 x_n\Omega_0$, 
where $m_n$ is the mass of the $n$-th graviton KK mode and the $x_n$ are the 
zeroes of the Bessel function $J_1$ ($x_1\sim 3.8$, $x_n\sim x_1 + \pi (n-1)$).
A useful relation following from these equations is:
\vspace*{-.1in}
\beq
m_n=x_n {m_0\over\mpl} {\lphi\over\sqrt 6}\,,\quad
\mbox{implying}\quad 
m_1=15.5 \gev\times \left({\mompl\over
    0.01}\right)\left(\frac{\lphi}{1\tev}\right)=380\gev
     \left({\mompl}\right)^{2/3}\left(\frac{\lnda}{1\tev}\right)\,.
\label{m1form}
\vspace*{-.1in}
\eeq
Given a value for $\lphi$, if $m_0/\mpl$ were known, then all the KK
masses would be determined and, therefore, our predictions for $a_J$
would be unique.  However, additional theoretical arguments are needed
to set $m_0$ independently of $b_0$ ($m_0b_0\simeq 70$ is required to
solve the hierarchy problem). As noted earlier, reliability of the RS
model requires values for $m_0/\mpl<1$, in which case the $n=1$ KK
graviton is always below the cutoff $\lbar$.

String theory
estimates strongly suggest $\mompl\ll 1$, typically $\lsim 0.01$~\cite{Davoudiasl:1999jd}.
When $m_0/\mplanck$ is small, Eq.~(\ref{m1form}) implies
that one is summing over a very
large number of KK excitations. As $m_0/\mplanck$
increases, the number summed over slowly decreases. We will 
address later the experimental constraints on $\lphi$ as a function of
$\mompl$. 

\begin{figure}[h!]
  \bce
  \includegraphics[width=8cm,angle=90]{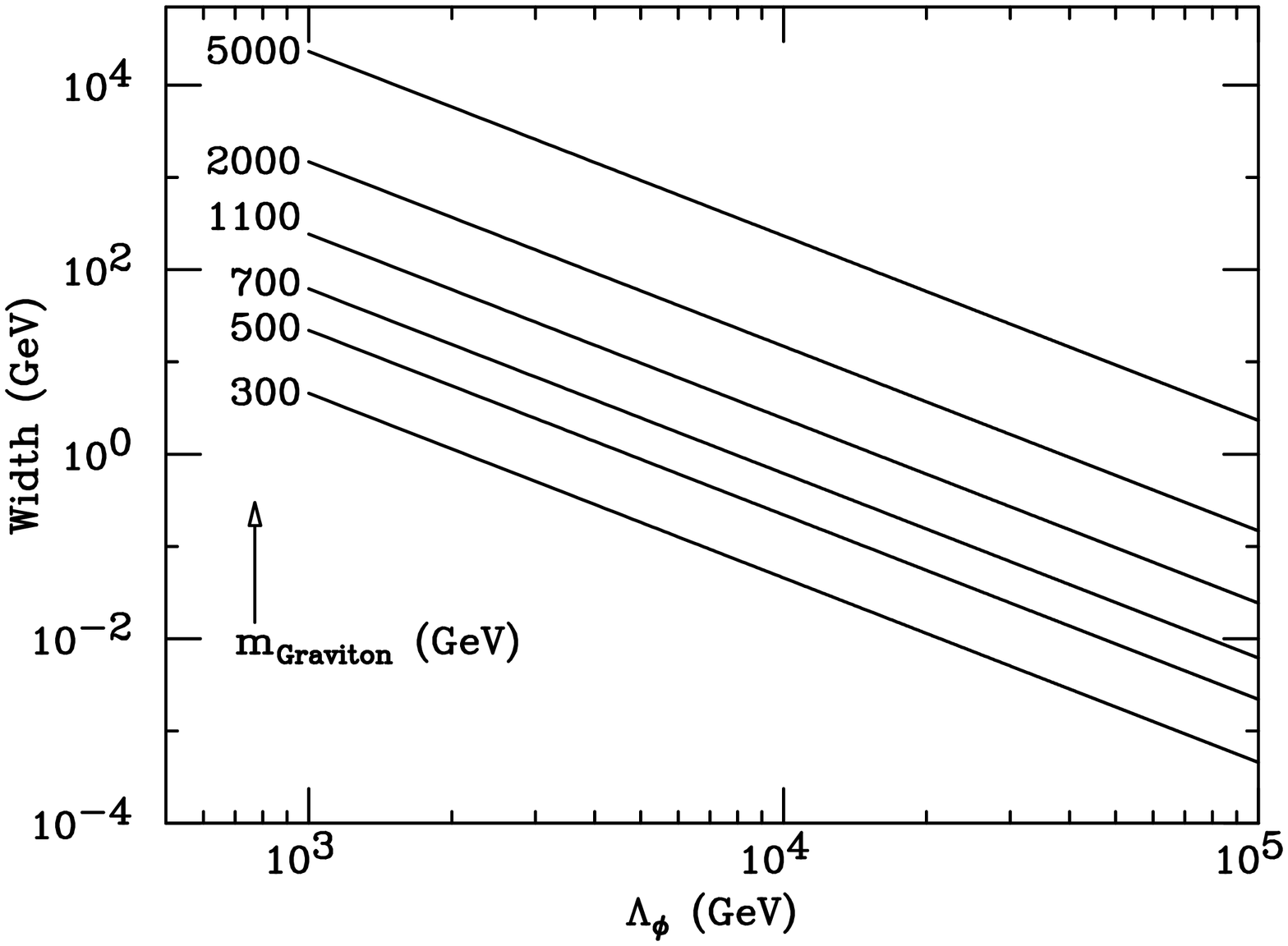}
\vspace*{-.3cm}
\caption{We plot the KK graviton width as a function of $\lphi$
for various values of the graviton mass.  This
plot applies independently of the level $n$ of the excitation. 
\label{grwidth}}
\ece
\vspace*{-.2in}
\end{figure}

We also need the widths of the gravitons:
\beq
\Gamma_n={6\over \lphi^2}\rho m_n^3\,,\quad\mbox{where}\quad 
 \rho={1\over 160\pi}\Sigma\,.
\label{gamform}
\eeq
In the above, $\Sigma$ counts the number of SM states to which $n$-th KK state
can decay.  For $m_n\gg 2\mz,2m_t$, $\Sigma=97/4$  with a possible
additional contribution of order $1/3$  coming from decays to Higgs
and radion states when accessible. Fig.~\ref{grwidth} shows the
graviton width for a number of cases. Note that it can be either very 
small (large $\lphi$, small mass) or very large (small $\lphi$, large
mass). The graviton widths are implicitly dependent upon $\mompl$.  
Indeed, Eq.~(\ref{gamform}) can be combined with Eq.~(\ref{m1form}) to
give 
\beq
\Gamma_n=\rho m_n x_n^2 \left({m_0\over
    \mpl}\right)^2\,,\quad\mbox{from which}\quad
{m_0\over \mpl}=\left[{\Gamma_n\over \rho m_n x_n^2}\right]^{1/2}\,.
\label{gamman}
\eeq 
Thus a measurement of $\Gamma_n$ and $m_n$ will determine $\mompl$
if $n$ is known.  

Current Tevatron constraints combined with the simplest implementation
of precision electroweak constraints assuming $\mh\lsim 350\gev$
jointly imply $m_1\gsim 650\gev$ for $0.01\leq \mompl\leq0.1$
\cite{magass}, which converts via Eq.~(\ref{m1form}) to
$(\mompl)(\lphi/1 \tev)\gsim 0.4$ or
$(\mompl)^{2/3}(\lnda/1\tev)>2.3$.  For example, if $\lphi=10\tev$,
the first equation gives $\mompl\gsim 0.04$.  However, the precision
electroweak constraints employed~\cite{Davoudiasl:2000wi} to obtain
the above limits assumed $\mh<350\gev$. We will be particularly
interested in cases where the Higgs mass is large.  In the absence of
Higgs-radion mixing, and neglecting KK gravitons, there are some
results at high $\mh$ from~\cite{Csaki:2000zn}.  There, it is noted
that there is a large uncertainty in the precision electroweak
calculations due to the non-renormalizable operators associated with
physics at and above the RS model cutoff scale.  They find that for
large non-renormalizable operators a very heavy Higgs can be
consistent with precision electroweak data even in the absence of
Higgs-radion mixing and neglecting KK contributions.  The result of
having both a heavy Higgs boson and KK contributions along with large
non-renormalizable operators is not known.  Thus, we believe it is
premature to use precision electroweak data to constrain the
parameters of the model.

In this case, the most reliable available constraints are those from
direct KK graviton production at the Tevatron. The most recent results
of which we are aware are those given in~\cite{magass}. These allow
much lower values of $m_1$ as compared to the values quoted in the
preceding paragraph, \eg\ $m_1>240\gev$ at $\mompl= 0.01$
(corresponding to $\lphi>15.5\tev$) rising to $m_1>700\gev$ at
$\mompl=0.05$ ($\lphi>9\tev$) and $m_1>865\gev$ at $\mompl =0.1$
($\lphi>5.58\tev$). As we shall see, it is unfortunate that no
Tevatron limits have been given for $\mompl<0.01$, a region that will
turn out to be of particular interest for us, and for which there is
some theoretical prejudice.  We are not certain if there are
additional experimental issues associated with detecting the very
light KK resonances (assuming moderate $\lphi$) that would be present.
We urge that the Tevatron analyses be extended into this region.
Weaker bounds derive from considering the 4-fermion effective
operators coming from exchange of massive KK
resonances~\cite{Davoudiasl:1999jd}. We are not certain how to extend
these $\mompl\geq 0.01$ results to cases where $\mompl$ is very small
and the lower lying KK gravitons are very light. However, naive
extrapolation of the plots in Fig.~3 of~\cite{Davoudiasl:1999jd}
suggest that only fairly large $\lphi$ values, $\gsim 20\tev$, might
not be excluded for very small $\mompl$.

\vspace*{-.07in}
\section{Numerical Results}
\label{numerics}
\vspace*{-.07in}

\begin{figure}[h]
  \bce
  \includegraphics[width=8cm,angle=90]{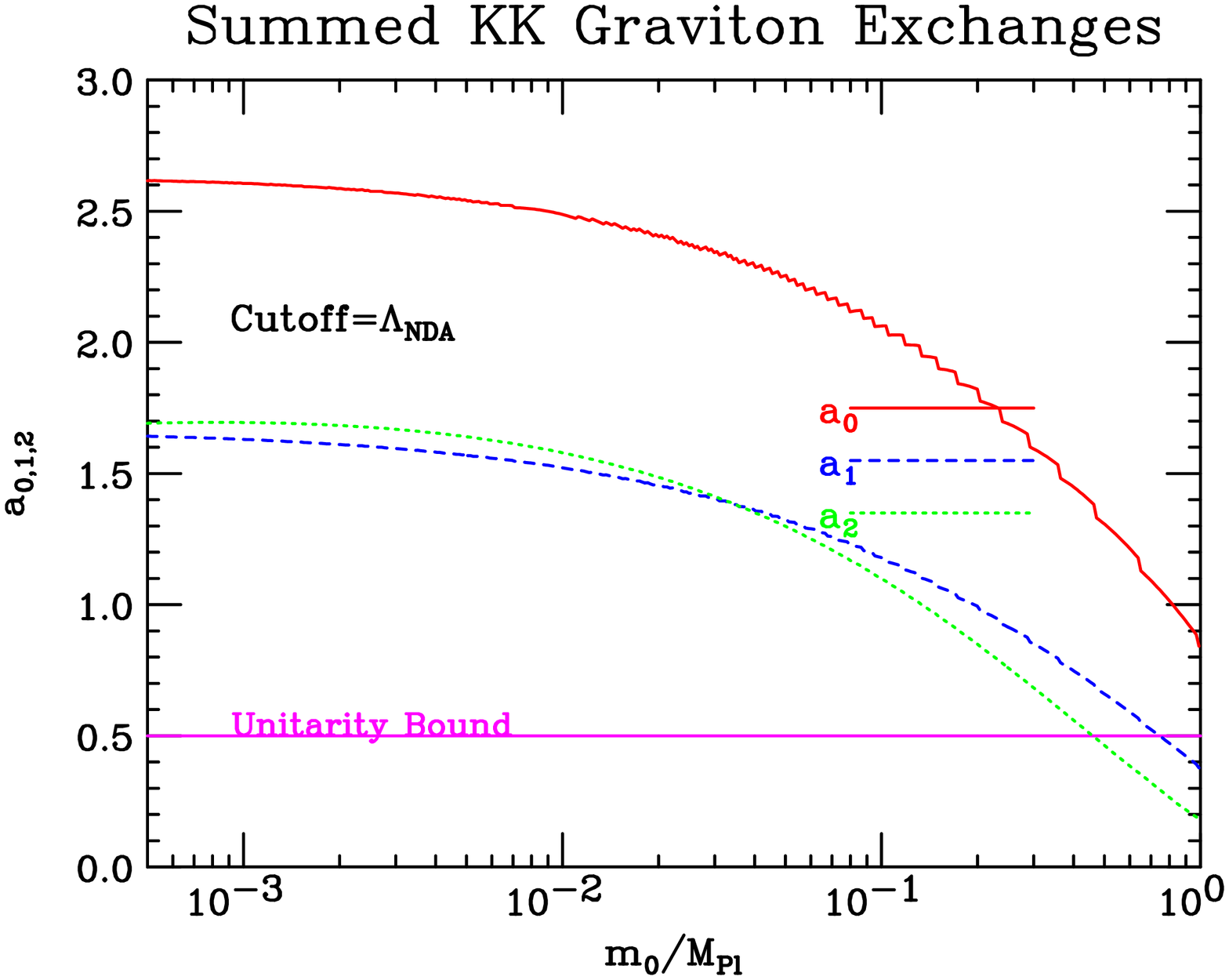}
  \vspace*{-.2cm}
\caption{We plot $\re a_{0,1,2}$ 
 as  functions of $\mompl$ as computed at $\rts=\lnda$ and summing
 over all KK graviton resonances with mass below $\lnda$, but without
 including Higgs or radion exchanges.
\label{lndaplot}}
\ece
\vspace*{-0.2in}
\end{figure}

We will now discuss in detail 
the importance of the KK gravitons in determining
unitarity constraints on the model.  We begin by presenting
in Fig.~\ref{lndaplot} the values of $\re a_{0,1,2}$ as functions
of $\mompl$ obtained by taking $\rts=\lnda$ (implying that a different 
cutoff is employed for each value of $m_0/\mpl$)  and summing over all KK
resonances with mass below $\lnda$.  
We see that $\wlwl$
scattering violates unitarity if $\lnda$ is employed as the cutoff.
A more appropriate cutoff is determined 
numerically by requiring $|\re a_{0,1,2}|<1/2$  for $\rts=\lbar$ after
summing over KK resonances with mass below $\lbar$.
\begin{figure}[h]
  \bce
  \includegraphics[width=6.8cm,angle=90]{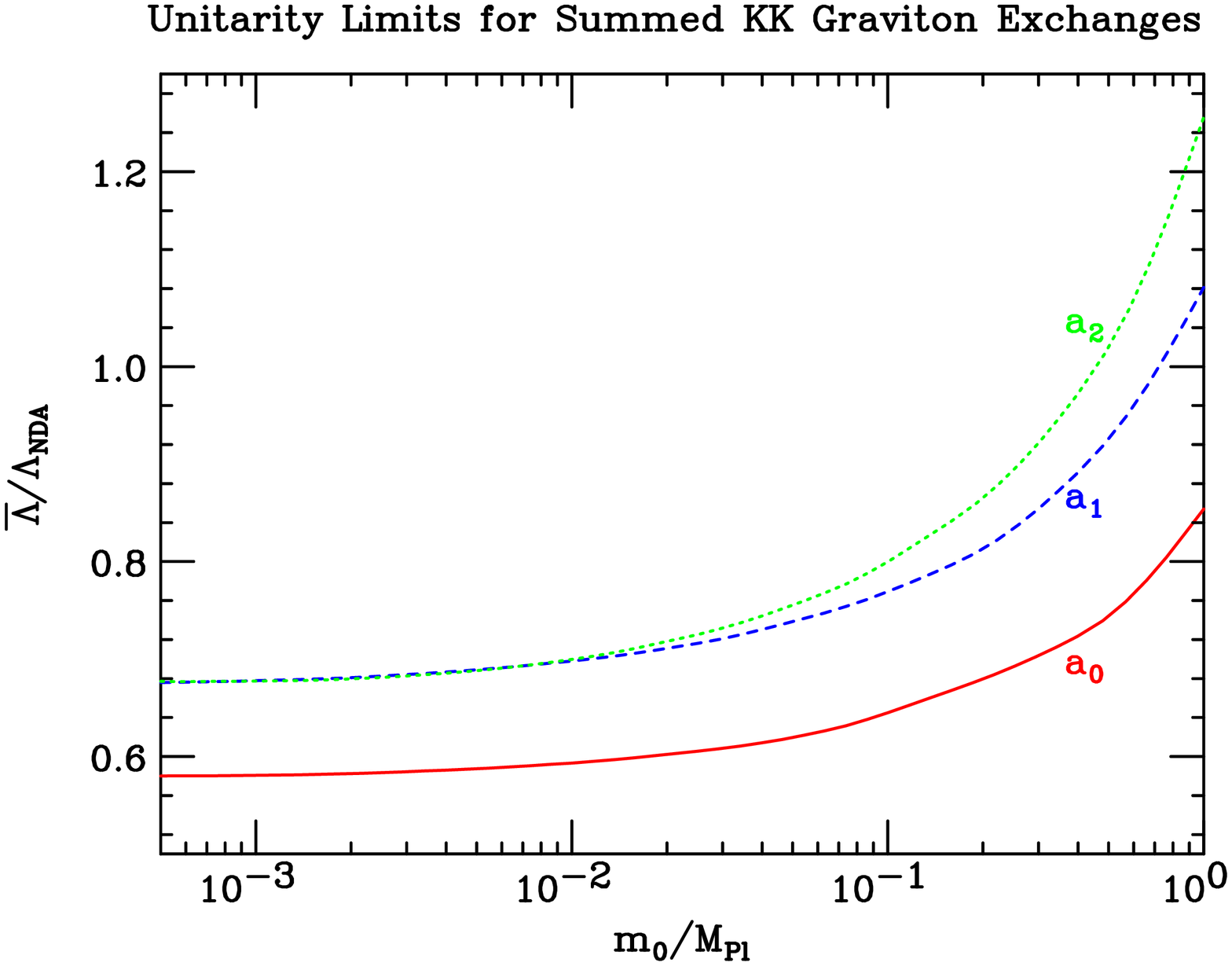}
  \includegraphics[width=6.8cm,angle=90]{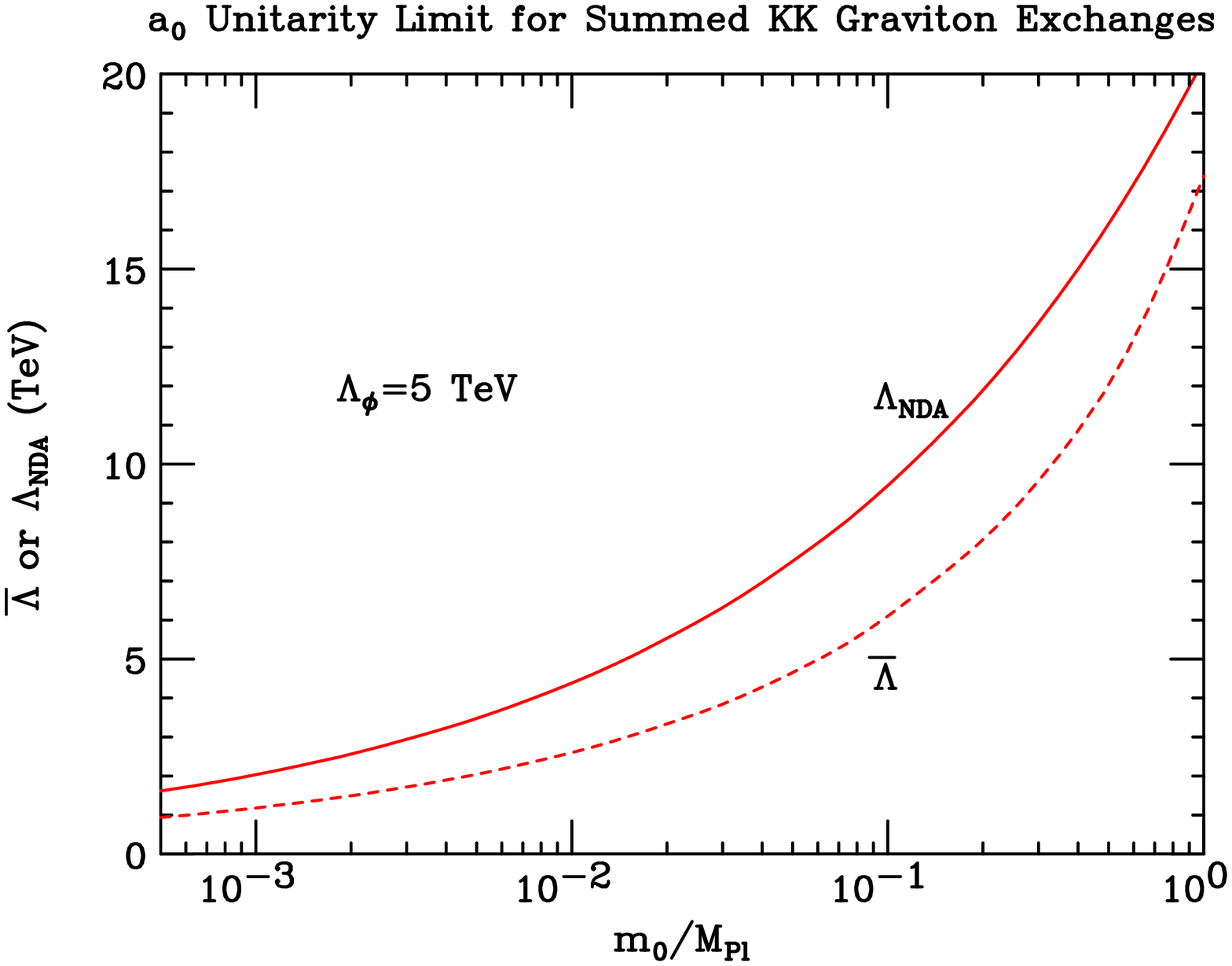}
  \vspace*{-.3cm}
\caption{In the left hand plot, we give $\lbar/\lnda$ as a function of $\mompl$, where
  $\lbar$ is the largest $\rts$ for which $\wlwl$ scattering is
  unitary after including KK graviton exchanges with mass up to
  $\lbar$, but before including Higgs and radion exchanges. Results
  are shown for the $J=0$, 1 and 2 partial waves. With increasing
  $\rts$ unitarity is always violated earliest in the $J=0$ partial
  wave, implying that $J=0$ yields the lowest $\lbar$. The right hand
  plot shows the individual absolute values of $\lbar(J=0)$ and
  $\lnda$ for the case of $\lphi=5\tev$; $\lbar/\lnda$ is independent
  of $\lphi$
\label{lndaratio}}
\ece
\vspace*{-0.2in}
\end{figure}
In the left-hand plot of Fig.~\ref{lndaratio}, we display the ratio
$\lbar/\lnda$ as a function of $\mompl$, where $\lbar$ is the largest
$\rts$ for which $\wlwl$ scattering is unitary when computed including
only the KK graviton exchanges. Results are shown for $J=0$, 1, and 2. As a
function of $\rts$, the $J=0$ partial wave is always the first to
violate unitarity and gives the lowest value of $\lbar$.  As discussed
in the introduction, we adopt the conservative approach of 
defining $\lbar$ to be the $J=0$ value. We will cut off
our sums over KK exchanges when the KK mass reaches this $J=0$ value.
We see that the $\lbar$ so defined is typically a significant fraction
of $\lnda$, but never as large as $\lnda$.  Still, it is quite
interesting that the unitarity consistency limit $\lbar$ tracks the
'naive' $\lnda$ estimate fairly well as $\mompl$ changes over a wide
range of values. (A very rough 'derivation' of this result
  appears in our brief Appendix on approximate derivations.)  The
right-hand plot of Fig.~\ref{lndaratio} shows the actual values of
$\lbar$ and $\lnda$ as functions of $\mompl$ for the case of
$\lphi=5\tev$. Note that for larger $\mompl$ they substantially exceed
the input inverse coupling scale $\lphi$, whereas for smaller $\mompl$
they are both substantially below $\lphi$. In other words, using
either $\lbar$ or $\lnda$, one concludes that $\lphi$, and equally
$\lwh$, are themselves not appropriate estimators for the maximum
scale of validity of the model.

\begin{figure}[h]
  \bce
  \includegraphics[width=7cm,angle=90]{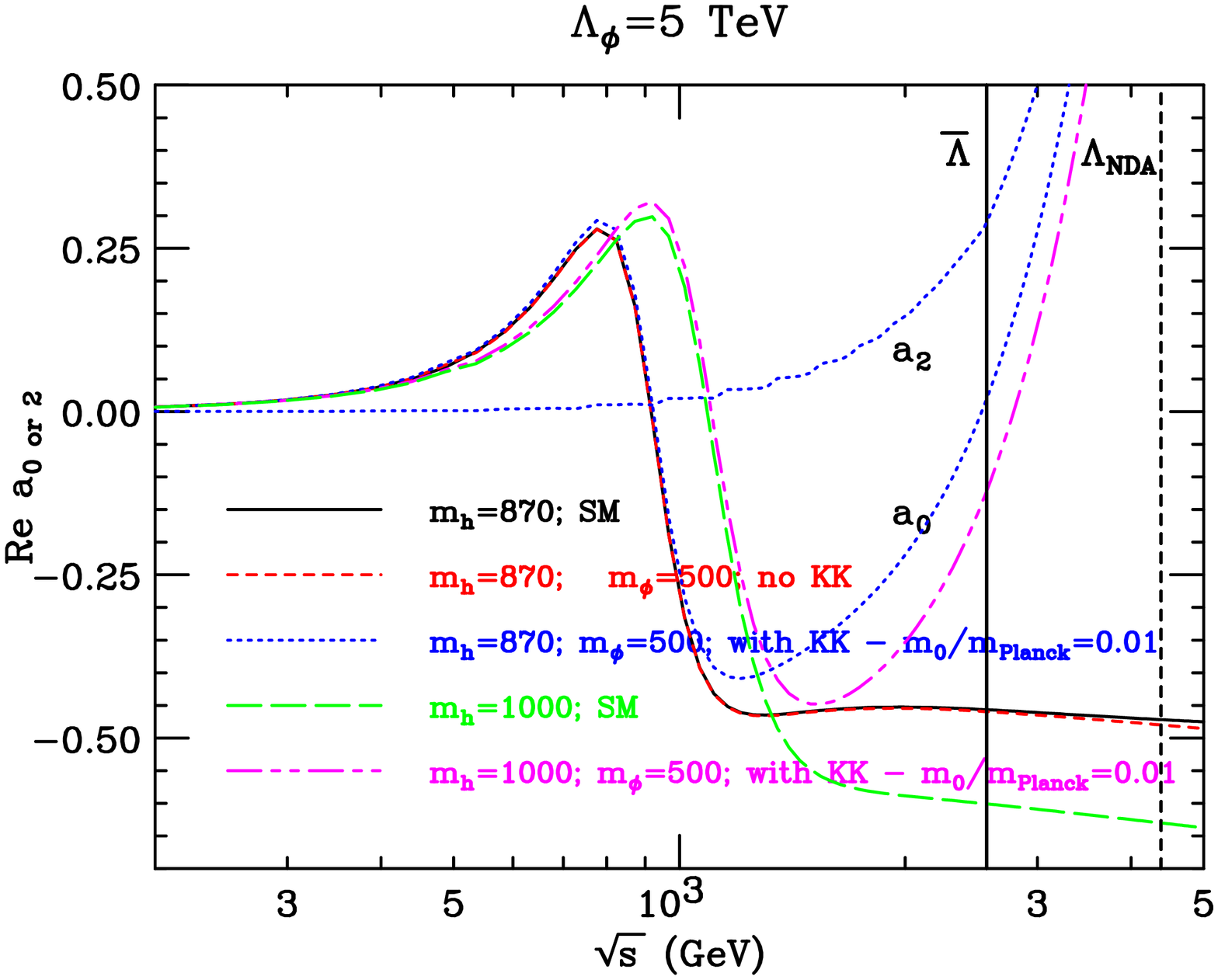}
  \includegraphics[width=7cm,angle=90]{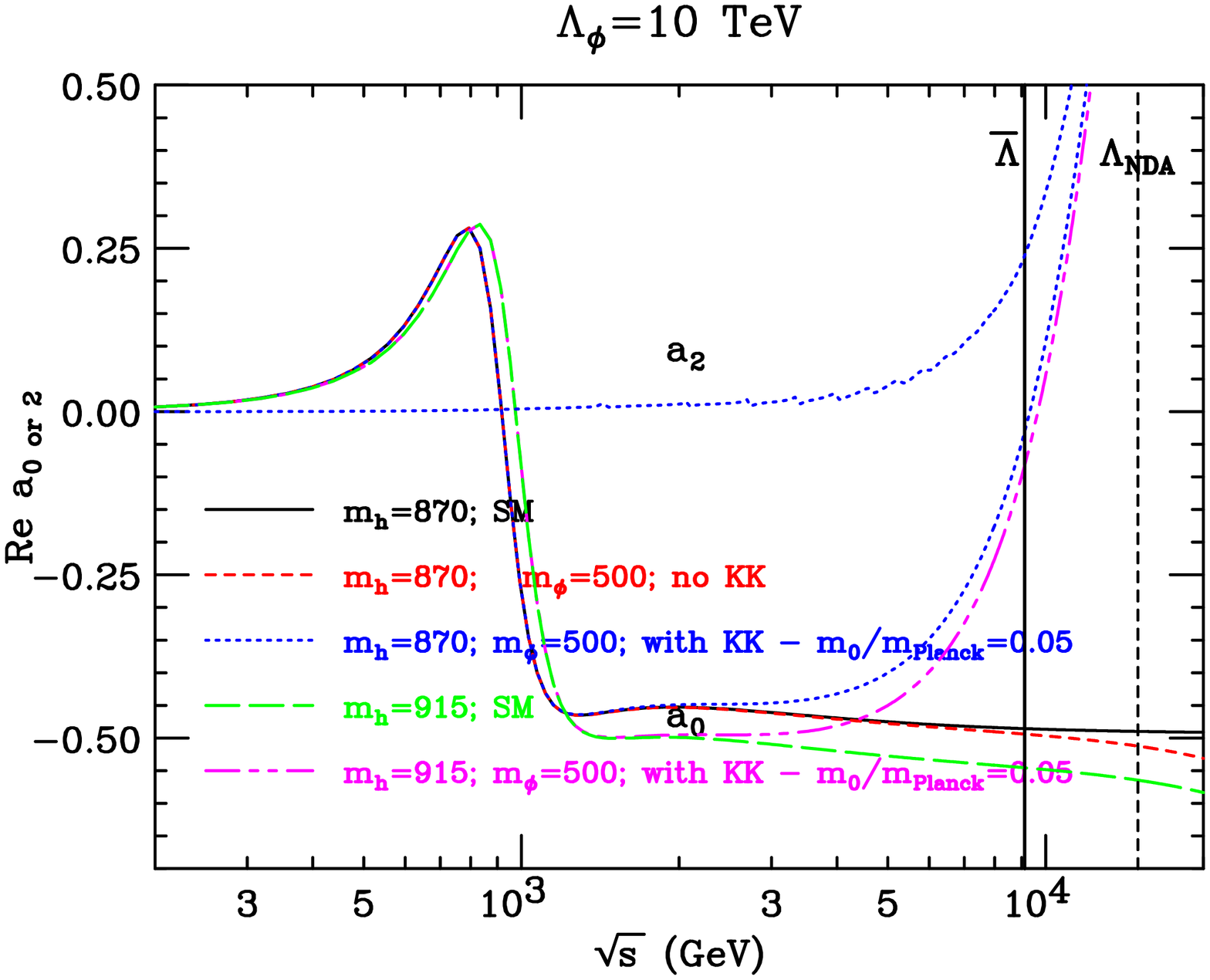}
  \vspace*{-.3cm}
\caption{For $\lphi=5\tev$ --- left ($\lphi=10\tev$ --- right), we plot $\re a_0$ as a function of $\rts$ 
for five cases: 1) solid (black) $\mh=870\gev$, SM contributions only
($\gamma=0$); 2) short dashes (red) $\mh=870\gev$, with an unmixed
 radion of mass $\mphi=500\gev$ included, but no KK gravitons (we do not show
the very narrow $\phi$ resonance);  3) dots (blue) as in 2), but including the sum
over KK gravitons taking $\mompl=0.01$ ($\mompl=0.05$) --- $\re a_2$ is also shown for this case; 4) long
dashes (green) $\mh=1000\gev$ ($915\gev$),
with an unmixed radion of mass $\mphi=500\gev$, but no KK gravitons);
5) as in 4), but including the sum over KK gravitons taking
$\mompl=0.01$ ($\mompl=0.05$). The $\lbar$ and 
$\lnda$ values for $\mompl=0.01$ ($\mompl=0.05$) are indicated by vertical lines. 
\label{svariationcases}}
\ece
\vspace*{-.1in}
\end{figure}

It is worth noticing that the results obtained here can also be understood 
in terms of the strategy developed in \cite{Giudice:1998ck}, where 
the discrete summation over KK gravitons is replaced by an integration
over a continuous mass distribution when $\mompl \ll 1$.
In that limit, our results for the (relatively unimportant) $s$-channel contributions can be reproduced
adopting the method of \cite{Giudice:1998ck}. However, it should be emphasized
that for moderate values of $\mompl$ the discrete summation employed
here is a more accurate procedure.

Let us now examine how the presence of a heavy Higgs boson affects
$\wlwl$ unitarity, both before and after inclusion of the exchanges of
KK gravitons with mass below $\lbar$ (where $\lbar$ is the $J=0$
cutoff shown in Fig.\ref{lndaratio}).  In the left-hand plot of
Fig.~\ref{svariationcases} we give $\re a_0$ as a function of $\rts$
for the case of $\lphi=5\tev$ for two different $\mh$ values and with
and without radion and/or KK gravitons included. In the case where we
include only the SM contributions for $\mh=870\gev$, the figure shows
that $\re a_0$ asymptotes to a negative value very close to $-1/2$,
implying that $\mh=870\gev$ is very near the largest value of $\mh$
for which unitarity is satisfied in $\wlwl$ scattering in the SM
context. If we add in just the radion contributions (for
$\mphi=500\gev$~\footnote{Because of the smallness of the $\gam^2$
  multiplying $\mphi^2$ in the expression for $\ms$, there is little
  change of our results as a function of $\mphi$ in the range
  $\mphi\in[10,1000]\gev$ so long as $\lphi$ is above $1\tev$.} -- the
$\phi$ resonance is very narrow and is not shown), then a sharp-eyed
reader will see (red dashes) that $\re a_0$ is a bit more negative at
the highest $\rts$ plotted, implying earlier violation of unitarity.
However, if we now include the full set of KK gravitons, which enter
with an increasingly positive contribution, taking $\mompl=0.01$
(dotted blue curve) one is far from violating unitarity due to $\re
a_0<-1/2$ for $\rts$ values above $\mh=870\gev$; instead, the positive
KK graviton contributions, which cure the unitarity problem at
negative $\re a_0$ for $\rts$ above $\mh$, cause unitarity to be
violated at large $\rts$, but {\it above} $\lbar$, as $\re a_0$ passes
through $+1/2$.  In fact, in the case of a heavy Higgs boson we see
that $\re a_2$ actually violates unitarity earlier than does $\re
a_0$. However, even using $\re a_2$ as the criterion, unitarity is
first violated for $\rts$ values above the $\lbar$ value appropriate
to the $\mompl=0.01$ value being considered, but still below $\lnda$.
In fact, it is very generally the case that unitarity is not violated
at $\rts=\lbar$ (which is typically a sizable fraction of $\lnda$) no
matter how small we take $\mompl$. However, as we shall see, unitarity
can be violated in the vicinity of $\rts\sim \mh$ if $\mh$ is large
and $\mompl$ is sufficiently small.

Looking again at the left plot of Fig.~\ref{svariationcases}, we
observe that if $\mh$ is increased to $1000\gev$, the purely SM plus
radion contributions (long green dashes) show strong unitarity
violation at large $\rts$ due to $\re a_0<-1/2$.  However, if we
include the KK gravitons (long dashes and two shorter dashes in
magenta), the negative $\re a_0$ unitarity violation disappears and
unitarity is instead violated at higher $\rts$.  Thus, it is the KK
gravitons that can easily control whether or not unitarity is violated
for $\rts<\lbar$ for a given value of $\mh$.

As discussed earlier, $\lphi=5\tev$ is actually too low a value for
consistency with Tevatron limits at $\mompl=0.01$.  Thus, in the right
hand plot of Fig.~\ref{svariationcases} we show $\re a_0$ and $\re
a_2$ for the case of $\lphi=10\tev$ and $\mompl=0.05$, a parameter set
that is allowed by Tevatron KK limits. The heavier Higgs mass is
chosen to be $915\gev$ in this case. The plot shows that $\mh=915\gev$
is just barely consistent with unitarity near $\re a_0=-1/2$ when KK gravitons
are included. If
$\lphi$ is increased further, keeping $\mompl=0.05$ fixed, the largest
$\mh$ consistent with avoiding unitarity violation at $\re a_0=-1/2$ 
will decrease towards the SM value of $\mh=870\gev$.

\begin{figure}[t!]
  \bce
  \includegraphics[width=7cm,angle=90]{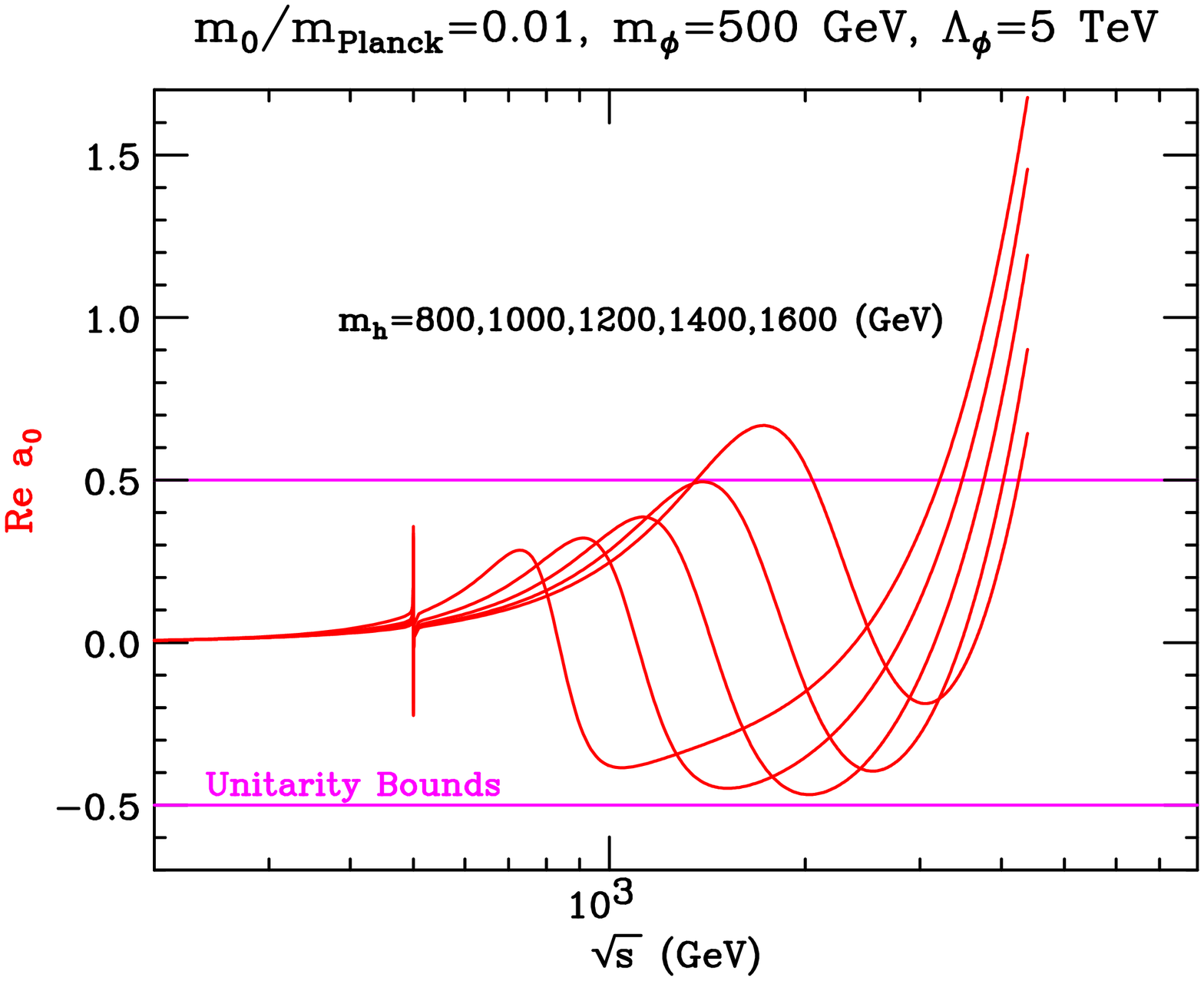}
  \includegraphics[width=7cm,angle=90]{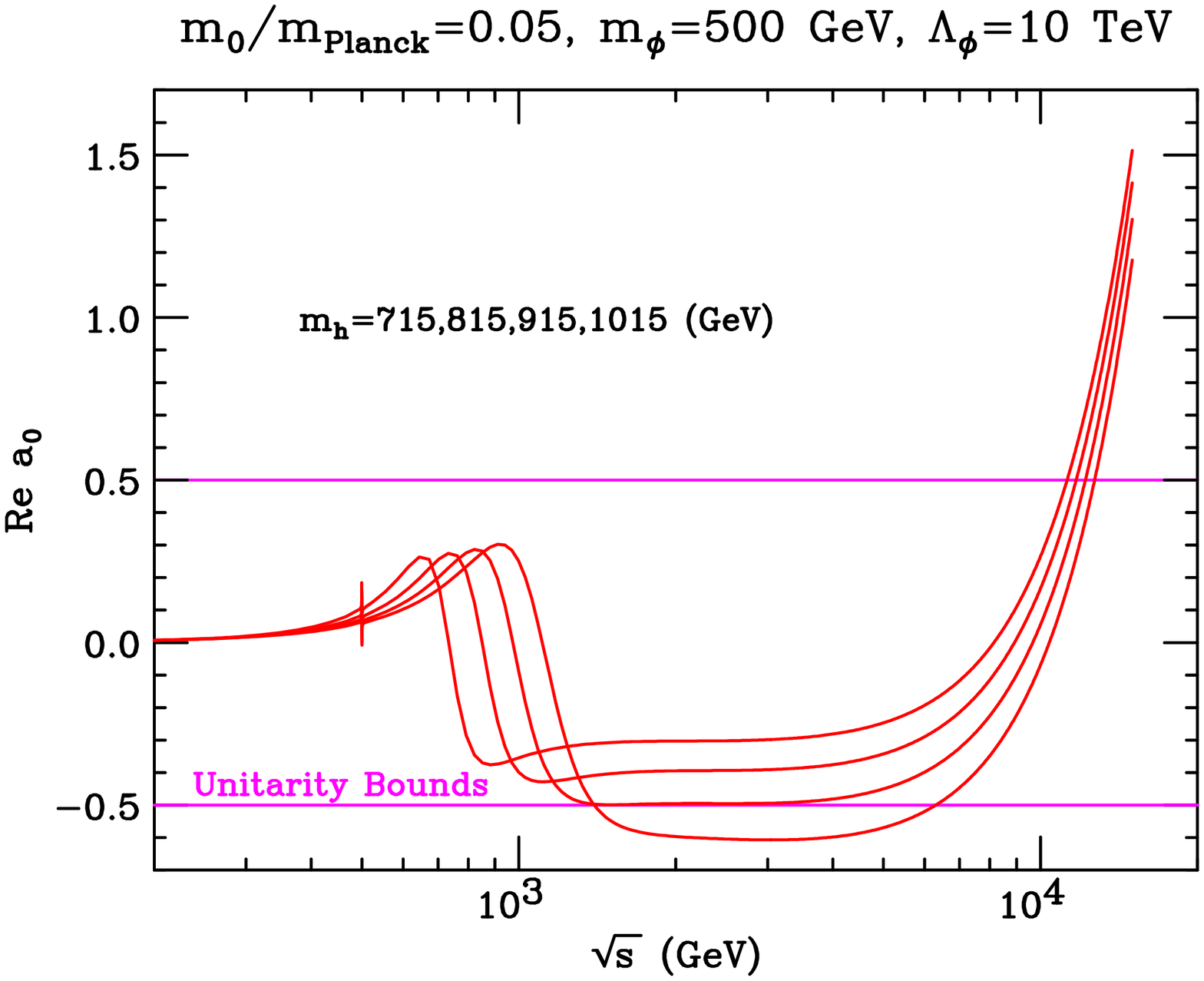}
  \vspace*{-.3cm}
\caption{We plot $\re a_0$ as a function of $\rts$ 
taking $\mompl=0.01$ and $\lphi=5\tev$ (left) or $\mompl=0.05$ and $\lphi=10\tev$ (right) for 
various Higgs masses. The curves terminate at
$\rts=\lnda$ as evaluated for a given $\mompl$.
\label{svariationmhcases}}
\ece
\vspace*{-.1in}
\end{figure}
In Fig.~\ref{svariationmhcases} we display more clearly variations
with $\mh$. The left plot shows that if $\lphi=5\tev$ and
$\mompl=0.01$, the positive KK graviton contributions guarantee that
$\re a_0$ never falls below $-1/2$ for any Higgs mass.  In fact, one
can increase $\mh$ to as high as $1400\gev$, at which point unitarity
violation occurs near $\rts=\mh$ at $\re a_0=+1/2$. The right plot of
Fig.~\ref{svariationmhcases} contrasts this with results for
$\lphi=10\tev$ and $\mompl=0.05$, a case consistent with Tevatron
limits on the lightest KK resonance.  For this case, the
maximum Higgs mass allowed by unitarity is $\mh=915\gev$ and is
determined by unitarity violation at $\re a_0=-1/2$.

\begin{figure}[h]
\bce
\includegraphics[width=7cm,angle=90]{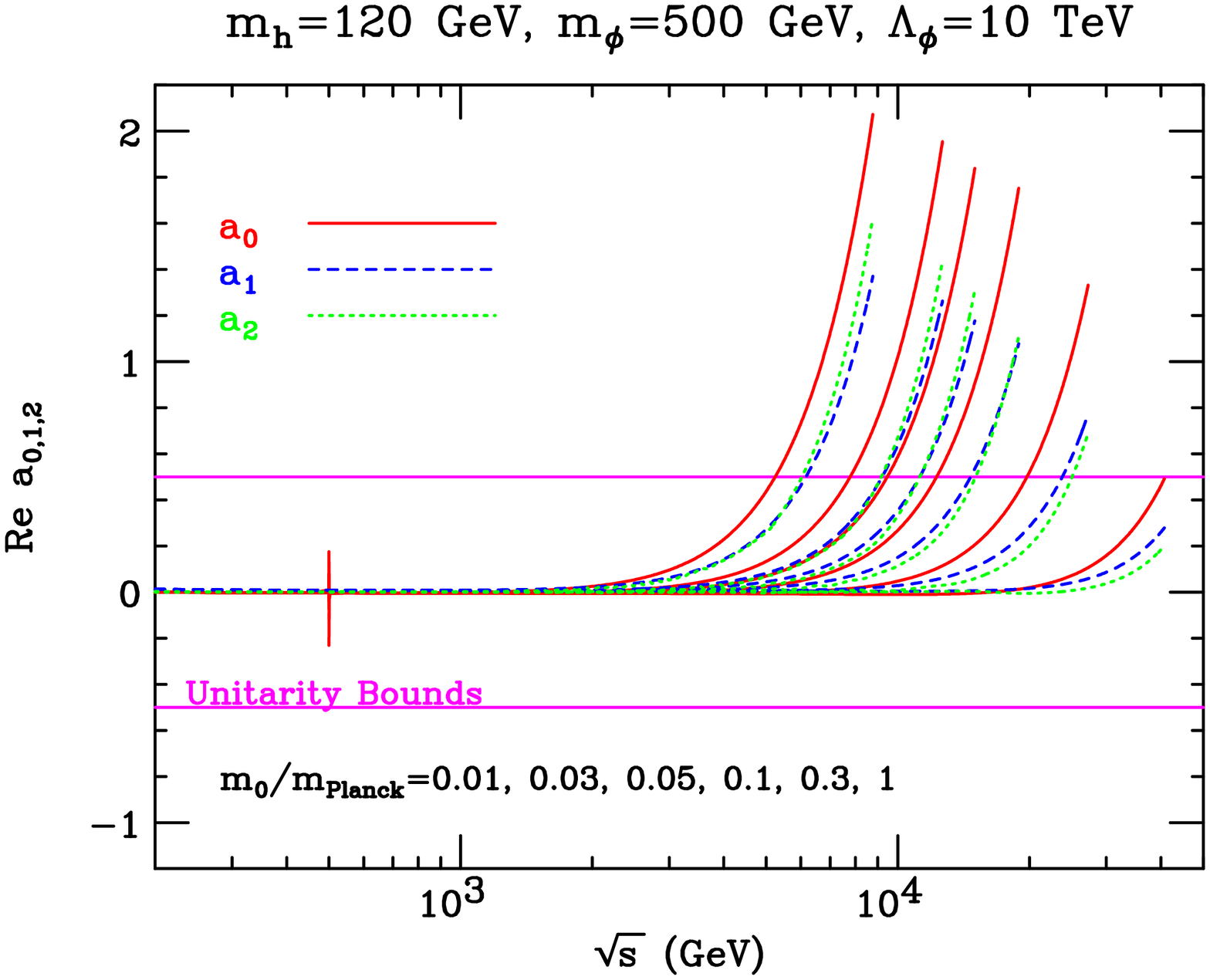}
\includegraphics[width=7cm,angle=90]{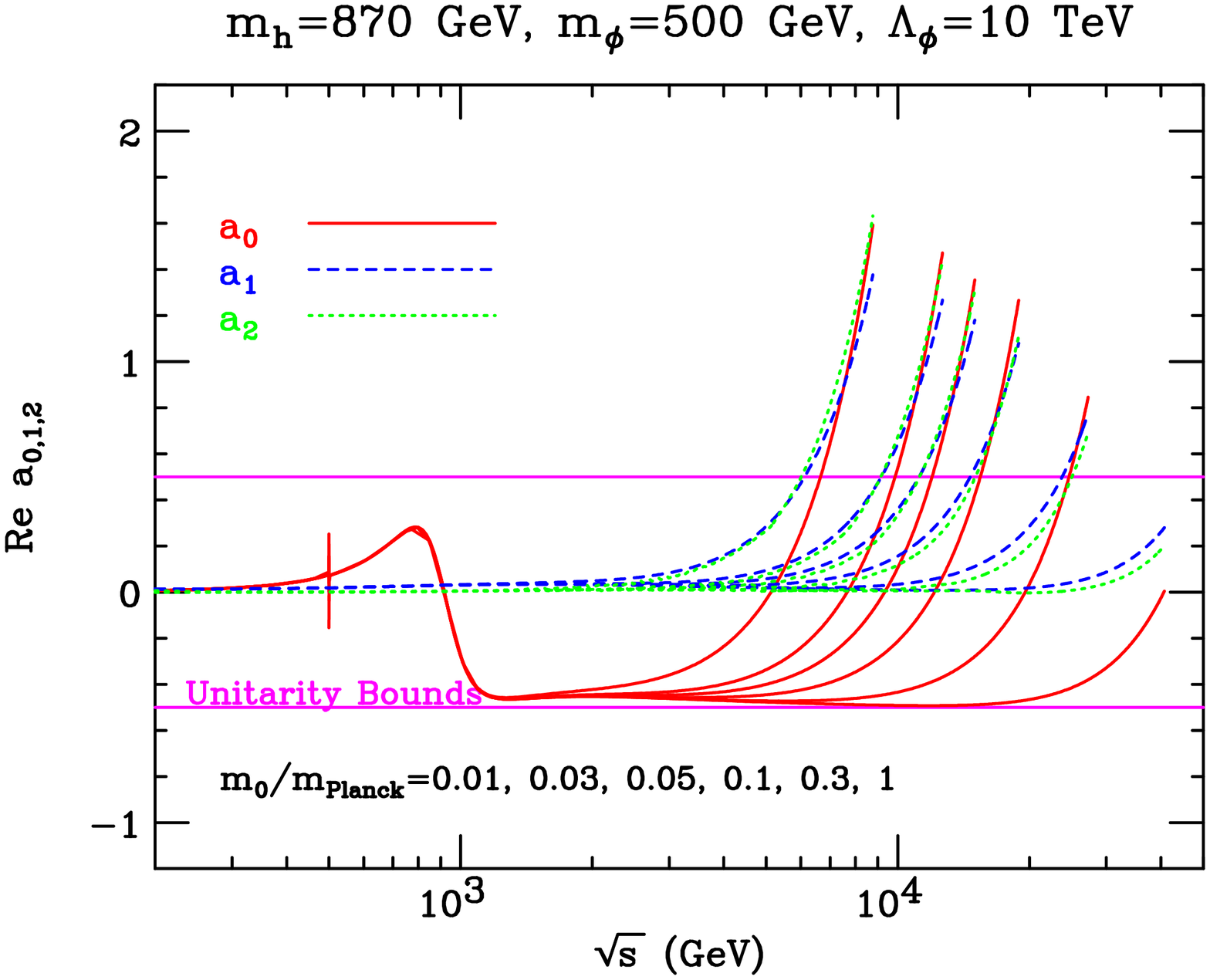}
\vspace*{-.3cm}
\caption{We plot $\re a_{0,1,2}$ as functions of $\rts$ for $\mh=120\gev$ and $\mh=870\gev$,
  taking
$\mphi=500\gev$ and $\lphi=10\tev$, and for the $\mompl$ values
indicated on the plot. Curves of a given type become higher
as  one moves to lower $\mompl$ values. 
We have included all KK resonances with $m_n<\lbar$ (at
all $\rts$ values). Each curve terminates at $\rts=\lnda$, where
$\lnda$ at a given $\mompl$ is as plotted earlier in
Fig.~\ref{lndaratio}. The value of $\rts$ at which a given curve
crosses above $\re a_0=+1/2$ is always slightly above the $\lbar$
(plotted in Fig.~\ref{lndaratio})
value for the given $\mompl$. \label{a012rts}}
\ece
\vspace*{-0.4cm}
\end{figure}

\begin{figure}[h]
\bce
\includegraphics[width=7cm,angle=90]{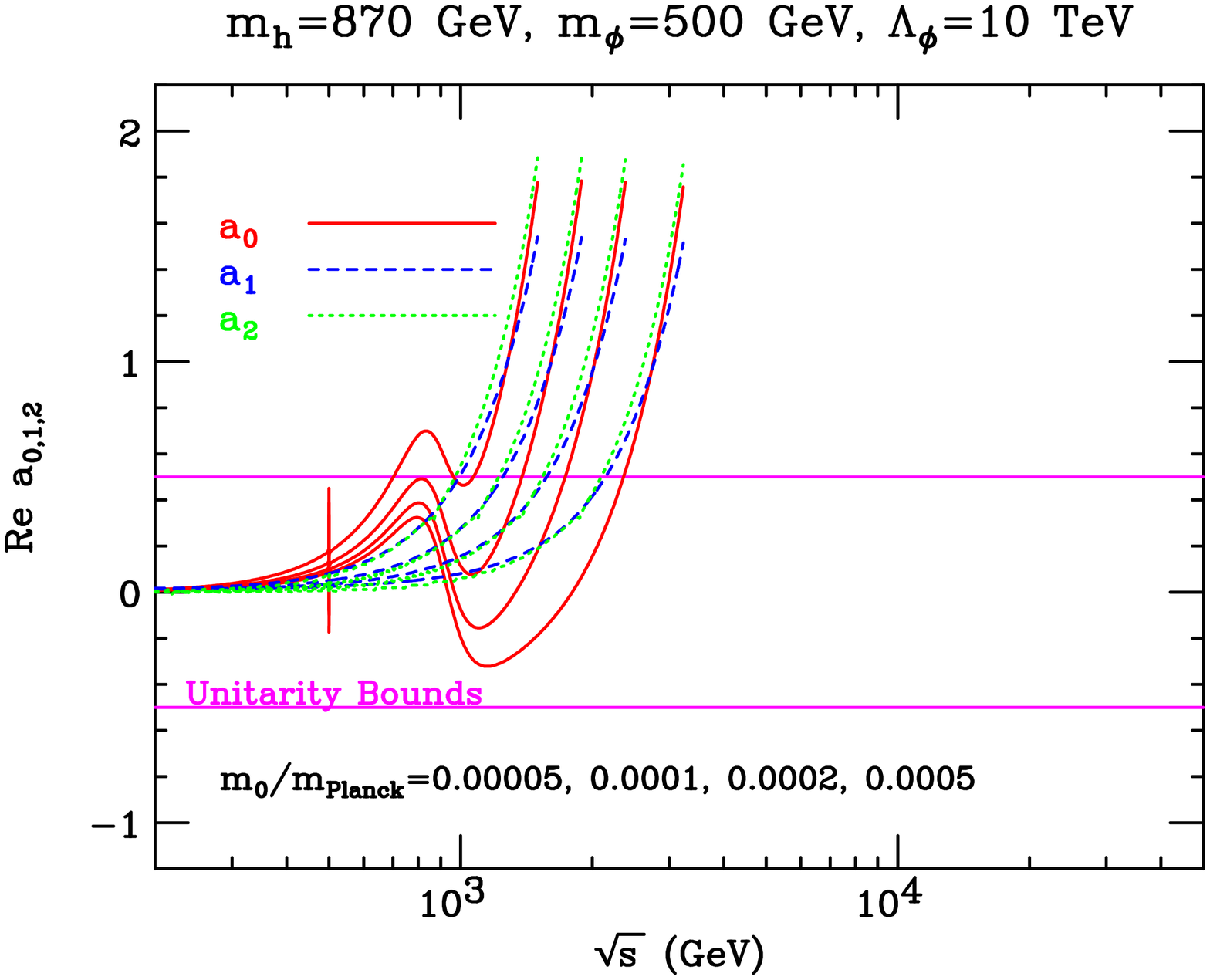}
\includegraphics[width=7cm,angle=90]{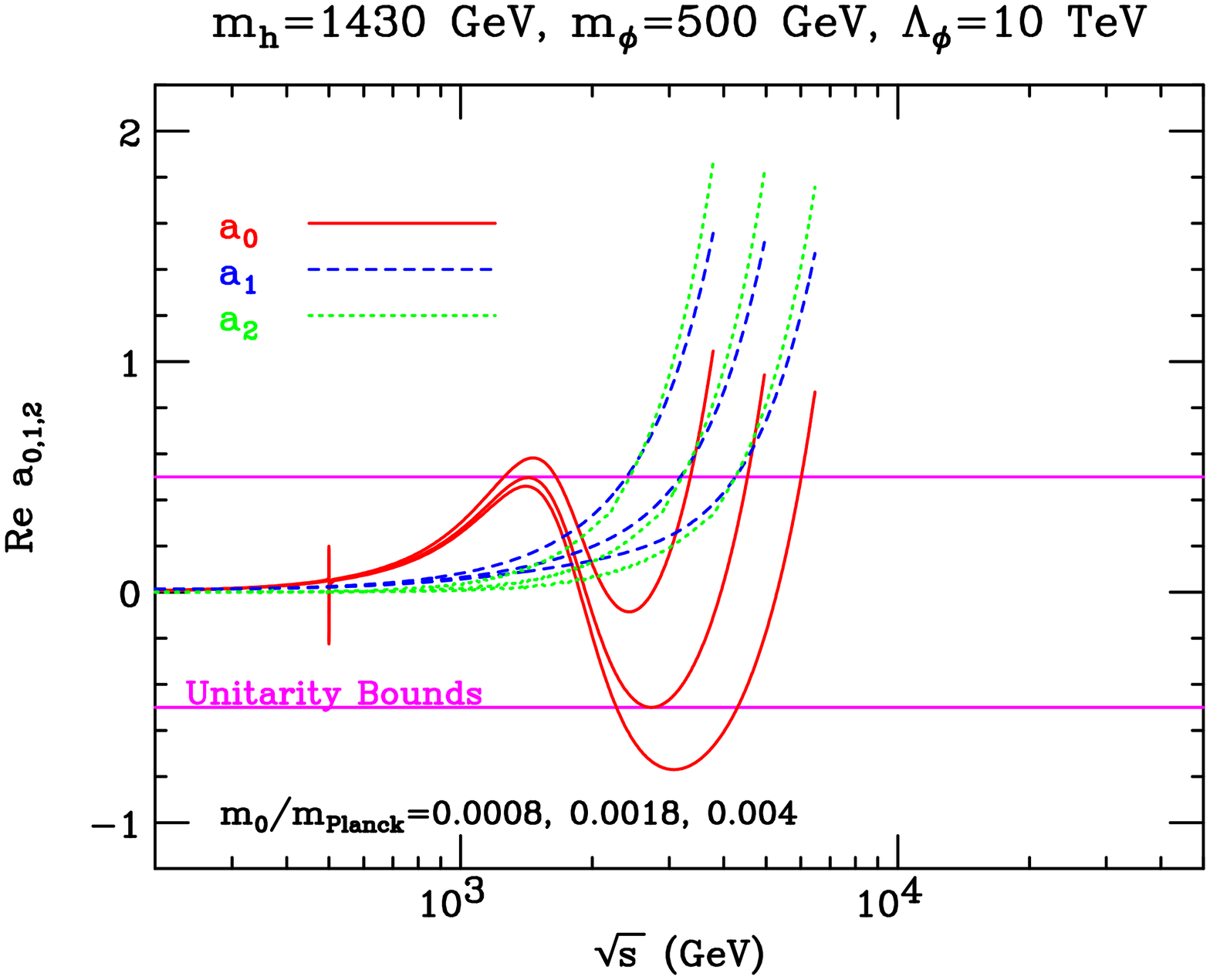}
\vspace*{-.3cm}
\caption{We plot $\re a_{0,1,2}$  as functions of $\rts$ for $\mh=870\gev$ and $\mh=1430\gev$,
  taking
$\mphi=500\gev$ and $\lphi=10\tev$, and 
for the $\mompl$ values
indicated on the plot. Otherwise, as in Fig.~\ref{a012rts}.\label{a012rts2}}
\ece
\vspace*{-0.6cm}
\end{figure}

Thus, the Higgs plus vector boson exchange contributions have a large
affect on the behavior for $\re a_0$ (whereas the radion exchange
contributions are typically quite small in comparison).  Let us
consider further cases of $\mh$ and $\mompl$ at fixed $\mphi=500\gev$,
taking $\lphi=10\tev$ so as to indicate the influence of the KK
resonance tower for an $\lphi$ choice that for some $\mompl$ values
can be consistent with the Tevatron limits summarized earlier.  We
first consider $\mh=120\gev$ and $\mh=870\gev$.  Fig.~\ref{a012rts}
shows the behavior of the real parts of the $a_{0,1,2}$ partial waves
as a function of $\sqrt s$ for a series of $\mompl$ values.  For
$\mh=120\gev$, the behavior of the amplitudes for $\rts>200\gev$ is
almost the same as in the absence of the Higgs and radion aside from
the (very narrow) radion resonance peak.  This is to be contrasted
with the case of $\mh=870\gev$, for which one
observes a (very broad) Higgs resonance in $\re a_0$, followed by a
strong rise (depending on $\mompl$) due to KK graviton exchanges. (For
$J=1$ and 2 there is no resonance structure associated with the Higgs
or radion since scalars do not contribute to $a_{1,2}$ in the
s-channel.) As we have already noted, in the vicinity of the possible
unitarity violation at negative $a_0$ from the SM plus radion
contributions, the KK graviton exchanges give a possibly very relevant
positive contribution to $\re a_0$.

A further plot for $\mh=870\gev$ and $\lphi=10\tev$, but focusing on
much smaller values of $\mompl$, appears as the left-hand plot of
Fig.~\ref{a012rts2}.  Note that for the very small value of
$\mompl=0.0001$, unitarity is only just satisfied for $\rts\sim \mh$
and that $\re a_0$ exceeds $+1/2$ near $\rts\sim \mh$ for
$\mompl<0.0001$. This is a general feature in the case of a heavy
Higgs; there is always a lower bound on $\mompl$ coming purely from
unitarity.  The
right-hand plot of Fig.~\ref{a012rts2} shows how high we can push the
mass of the Higgs boson without violating unitarity. For
$\mh=1430\gev$, we are just barely consistent with the unitarity limit
$|\re a_0|\leq 1/2$ (until large $\rts\gsim \lbar$) if $\mompl=0.0018$
(and $\lphi=10\tev$).  Any lower value of $\mompl$ leads to $\re
a_0>+1/2$ at $\rts\sim\mh$ and any higher value leads to an excursion
to $\re a_0<-1/2$ at higher $\rts$ values (but still below $\lbar$).
As discussed earlier, there are no limits of which we are aware
on the $\mompl$ values considered in Fig.~\ref{a012rts2} coming from
direct production of KK gravitons. For such values, the KK gravitons
would have very small masses. An analysis is needed.

\begin{figure}[h!]
\bce
\includegraphics[width=8cm,angle=90]{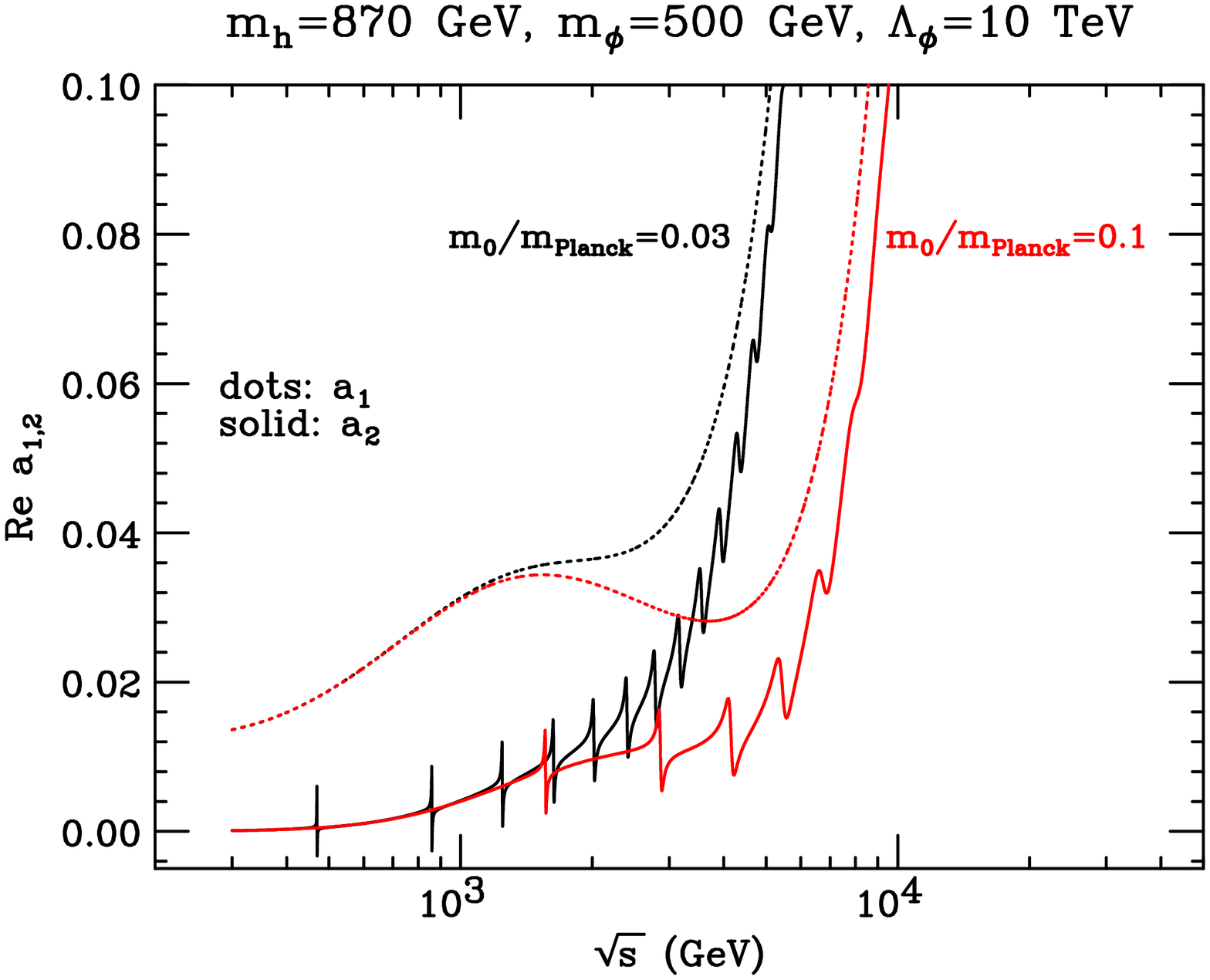}
\vspace*{-.2cm}
\caption{We plot $\re a_{1,2}$  for $\mh=870\gev$,
$\mphi=500\gev$ and $\lphi=10\tev$ 
as  functions of $\rts$ for the $\mompl$ values
indicated on the plot. These are the same curves as 
the corresponding ones in Fig.~\ref{a012rts}, but plotted on an
expanded scale so as to reveal the KK graviton resonances.
\label{gravresonances}}
\ece
\vspace*{-0.4cm}
\end{figure}

In order to actually see the graviton excitations in $a_2$ requires a
much finer scale for the plot. We illustrate for the case of
$\mh=870\gev$ and $\lphi=10\tev$ in Fig.~\ref{gravresonances}.  The
small KK resonance 
structures are apparent.  As seen from the figure, the resonant
graviton (spin 2) behavior is present only for $a_2$: no KK resonances
appear in $a_1$.~\footnote{There is also no graviton-resonance
  contributions to $a_0$. This fact serves as a non-trivial test of
  the calculation, since massive on-shell gravitons do not contain a
  $J=0$ component, see e.g.~\cite{Dicus:2004rt}.}  The $a_2$ KK resonance
peaks are suppressed by the partial $WW$ width to total width ratio.
We have not attempted to determine if the resonances could actually be
seen in $\wlwl$ scattering at the LHC or a future ILC.
However, it is important to note that other
authors~\cite{Davoudiasl:2000wi} have shown that the resonance masses,
cross sections and possibly widths can be measured in Drell-Yan
production for instance at the LHC and in many different modes at a
future ILC.

\begin{table}[h]
\begin{tabular}{|c|c|c|c|c|}
\hline
$\lphi(\tev)$ &  $5$ & $10$ & $20$ & $40$ \\
\hline\hline
\multicolumn{5}{|c|} {Absolute maximum Higgs mass} \\
\hline
$\mh^{\rm max}(\gev)$ & 1435 & 1430  & 1430 & 1430 \\
required $\mompl$ & $1.32\times 10^{-2}$ & $1.8\times 10^{-3}$ & $2.3\times 10^{-4}$ & $2.9\times
10^{-5}$ \\
associated $m_1(\gev)$ & 103.2 & 28.2 & 7.2 & 1.8 \\ 
\hline
\hline
\multicolumn{5}{|c|}{$\mompl=0.005$: Tevatron limit: $m_1>??$}\\
\hline 
$\mh^{\rm max}(\gev)$ & 1300 & 930 & 920 & 905 \\
associated $m_1(\gev)$ & 39 & 78 & 156 & 313 \\
\hline
\hline
\multicolumn{5}{|c|}{$\mompl=0.01$: Tevatron limit: $m_1>240\gev$}\\
\hline 
$\mh^{\rm max}(\gev)$ & 1405 & 930 & 910 & 895 \\
associated $m_1(\gev)$ & 78 & 156 & 313 & 626 \\
\hline\hline
\multicolumn{5}{|c|}{$\mompl=0.05$: Tevatron limit: $m_1>700\gev$}\\
\hline
$\mh^{\rm max}(\gev)$ & 930 & 915 & 900 & 885 \\
associated $m_1(\gev)$ & 391 & 782 & 1564  & 3129 \\
\hline\hline
\multicolumn{5}{|c|}{$\mompl=0.1$: Tevatron limit: $m_1>865\gev$}\\
\hline
$\mh^{\rm max}(\gev)$ & 920 & 910 & 893 & 883 \\
associated $m_1(\gev)$ & 782 & 1564 & 3128  & 6257 \\
\hline
\end{tabular}
\caption{Unitarity limits on $\mh$ for various $\lphi$ and $\mompl$
values. }
\label{primary}
\end{table}

In Table~\ref{primary}, we summarize the primary implications of our
results by showing a number of limits on $\mh$ for the choices of
$\lphi=5$, 10, 20 and $40\tev$.  The first block gives the very
largest $\mh$ that can be achieved, $\mhmax$, without violating
unitarity in $\wlwl$ scattering, along with the associated $\mompl$
value and mass $m_1$ of the lightest KK graviton.  Unfortunately, no
Tevatron limits have been given for the associated very small $\mompl$
values.  Even if they end up being experimentally excluded, it is
still interesting from a theoretical perspective that in the RS model
unitarity can be satisfied for all $\rts$ values below the $\lbar$
cutoff of the theory for a Higgs boson mass substantially higher than
the usual $870\gev$ value applicable in the SM context.  One finds
that $\mhmax$ is typically of order $1.4\tev$ if one chooses the
optimal value for $\mompl$. (A rough derivation of the value of
$\mhmax$ and associated optimal $\mompl$ is given in the brief
Appendix.)  It is also noteworthy that the required values of $\mompl$
are quite consistent with model expectations.

Table~\ref{primary} also gives the $\mhmax$ value achievable for the
four $\lphi$ cases listed above for various fixed $\mompl$.  Also
given are the associated $m_1$ values and the Tevatron direct
production limit when available.  For some of the cases that are
clearly consistent with Tevatron limits, unitarity is satisfied for
$\mh$ values as high as $\sim 915\gev$.  

Finally, we have seen that for a heavy enough Higgs boson mass, there
will be a value of $\mompl$ below which unitarity will be violated due
to $\re a_0>+1/2$ for $\rts\sim \mh$.  Even for the relatively large
value of $\mh\sim 870\gev$, the boundary value of $\mompl$ for
$\lphi=10\tev$ is already very small, $\mompl\sim 0.0001$, and this
boundary value decreases rapidly as $\mh$ is decreased. Such small
values are not favored in typical models and might also eventually be
ruled out by an appropriate analysis of Tevatron data designed to
exclude narrow KK resonances at very low mass.

\vspace*{-.07in}
\section{Parameter Determination from Experiment}
\label{experiment}
\vspace*{-.07in}

In this section we summarize how our results might be applicable when future LHC data
becomes available. As we have already noted, existing constraints
are important in assessing the relevant range of parameters to which
we should apply our results.

If the RS model is nature's choice, then the LHC and/or ILC can
potentially discover one or more graviton KK states. Their masses,
cross sections and widths will provide a lot of information, and, as
we shall discuss below, can strongly constrain the model. We will
focus on the means for determining $\mompl$ and $\lphi$. (Given these,
a determination of the cutoff $\lbar$ is possible.) To do so, it is
sufficient to determine $m_1$ and its coupling to SM particles, which,
see Eq.~(\ref{int}), is proportional to $1/\lwh=\sqrt 3/\lphi$. If
$m_1$ and $\lphi$ are known, then it is possible to determine $\mompl$
using Eq.~(\ref{m1form}).

If the graviton mass is known, then a measurement of the graviton width 
is one possible way to determine $\lphi$. 
This was illustrated
[after including phase space and non-asymptotic terms in
Eq.~(\ref{gamform})] in Fig.~\ref{grwidth} for a selection of possible
graviton masses in the range potentially observable at the LHC and/or
ILC.  Note that if the observed graviton is light, $m_G\lsim 200\gev$,
or $\lphi$ is large, the graviton width(s) are very small compared to
expected resolutions and cannot be used to extract $\lphi$ --- only a
lower bound on $\lphi$ could be extracted.  Once, the mass and width
are known, $\mompl$ can be determined from Eq.~(\ref{gamman}) if we
know which excitation level $n$ the KK resonance corresponds to.  

If resolving the graviton resonance shape or determining the excitation index
proves problematical, we should consider whether the absolute
magnitude of the cross
section for graviton production is a useful input. Consider first
$\epem \to G_n \to \mupmum$. It is easy to demonstrate that the peak
cross section at $s=m_n^2$ depends only on $m_n$ and not separately on
$\lphi$. Only if one can measure the shape of the cross section in the
vicinity of the peak can one obtain the width and thereby determine
$\lphi$.  However, as discussed above and shown in Fig.~\ref{grwidth},
for a large section of parameter space where $\lphi$ is moderate in
size and $\mompl$ is in the preferred range of $\sim 0.01$, the
graviton width will be much less than a GeV and a) a very fine scan
will be needed to even find the graviton and b) sufficiently fine scan
steps may not be possible to actually map out the shape of the
excitation. It is easier to extract $\lphi$ at fixed $m_n$ from the
hadron collider cross section in some given final state, which cross
section is proportional to $1/\lphi^2$ at fixed $m_n$. Useful plots
for $n=1$ appear in \cite{Allanach:2002gn} (see their Figs.~1, 6--8
and 10--11).
For $\mompl=0.01$ (0.05), the $\gam\gam$ final state provides a highly
accurate determination of $m_1$ and 20\% accuracy
or better for $\sigma(pp\to G_1) \br(G_1\to\gam\gam)$ for $m_G<1500\gev$ ($<3000\gev$).  

Thus, for a wide range of parameters we will be able to determine both
$\lphi$ and $\mompl$ with reasonable accuracy once LHC data is
available.  Given the measured values, we will be able to determine
the maximum Higgs mass for which unitarity remains valid for all
$\rts$ below the cutoff $\lbar$.  This might be useful if the Higgs
boson turns out to be significantly heavier than the SM limit derived
from unitarity, and therefore not easily seen as a clear resonance
structure.  Alternatively, if we find a Higgs boson and measure its
mass, we will be able to determine a fairly precise value for the
energy at which unitarity is violated and graviton interactions at the
loop level start to become strong.  This scale can be larger than
$\lbar$ if $\mh$ is large and would be the scale at which additional
new physics {\it must} enter.

%
\vspace*{-.1in}
\section{Summary and Conclusions}
\label{sum}
\vspace*{-.1in}

We have discussed perturbative unitarity for $\wlwl$ within the
Randall-Sundrum theory with two 3-branes and shown 
that the exchange of massive 4D Kaluza-Klein gravitons leads to
amplitudes growing linearly with the CM energy squared.  We have found
that the gravitational contributions cause a violation of unitarity
for $\rts$ below the natural cutoff of the theory, $\lnda$, as
estimated using naive dimensional analysis.  We have denoted by
$\lbar$ the maximum $\rts$ such that unitarity is still obeyed when
summing graviton exchange contributions for gravitons with mass below
$\lbar$.  Although $\lbar<\lnda$, there is a rough tracking of $\lbar$
and $\lnda$ in that the ratio $\lbar/\lnda$ ranges from $\sim 0.6$ for
small $\mompl$ to $\sim 0.85$ at $\mompl=1$, where $m_0$ is the
curvature of the RS metric and $\mpl$ is the usual 4d reduced Planck mass.
This means that for a wide range of $\mompl$ the two criteria are
roughly in agreement as to the maximum energy scale for which the RS
model will be a valid effective theory.

As we have shown, the KK resonance exchanges substantially modify
constraints from unitarity and these modifications can have very
important experimental implications.  To determine these implications,
consistent with the above discussion we sum over all KK gravitons with
mass below $\lbar$, regardless of the $\rts$ being considered. First,
it is important to note that the two basic RS model parameters $\lphi$
($\lphi^{-1}$ sets the strength of the couplings of matter fields to
KK resonances) and $\mompl$ can be extracted from experiment,
especially LHC observations of the first KK excitation.  If the Higgs
mass has also been measured, then the maximum $\rts$ for which $\wlwl$
scattering obeys unitarity in the RS model can be determined from the
results of this paper.  For modest $\mh$, this $\rts$ lies below
$\lnda$, but above $\lbar$.  More significantly, however, we have seen
that values of the Higgs mass much larger (up to $1.435\tev$ in the
sample case of $\lphi=5\tev$) than the usual $\mh\sim 870\gev$ SM
limit can be consistent with unitarity in $\wlwl$ scattering if the
$\mompl$ parameter is chosen below some (typically small for large
$\lphi$) value.  This happens by virtue of the fact that in the $\rts
\gsim \mh$ region, where $\re a_0$ can fall below $-1/2$ in the SM
context when $\mh$ is large, the KK graviton exchanges enter with a
positive sign and for small enough $\mompl$ there are enough
contributing KK exchanges that $\re a_0$ remains above $-1/2$.  In
such a case, unitarity is typically first violated by $\re a_1$ rising
above $+1/2$ at a $\rts$ value somewhat above $\lbar$.  In addition,
there is always a lower bound on $\mompl$ coming from the requirement
that $\re a_0$ not exceed $+1/2$ near the $\rts\sim \mh$ resonance
peak.  For $\mh\leq 870\gev$, and quoting results for $\lphi=10\tev$,
the $\mompl$ lower bound is very small (${\rm min}[\mompl]\leq 0.0001$) and not
particularly consistent with model expectations.  However, as $\mh$
increases towards $1.43\tev$ the lower bound on $\mompl$ increases
rapidly, until it reaches a reasonably model-friendly value of
$\mompl\sim 0.0018$ at $\mh\sim 1.43\tev$.  Experimental constraints
hint that a value as low as $\lphi=10\tev$ could be excluded when
$\mompl$ is small by a targeted analysis of Tevatron data; but
currently there are no constraints in this $\mompl$ region.  
For values of $\mompl$ that are clearly consistent with current Tevatron limits
we can only raise $\mh$ to $\sim
930\gev$ before violating unitarity in $\wlwl$ scattering.

Still stronger constraints from unitarity per se can be obtained if
one considers the full set of coupled channels ($WW$, $ZZ$, $hh$,
\ldots).  The complete approach would undoubtedly result in a somewhat
smaller value of $\lbar$ at a given $\mompl$. We have chosen to adopt
a somewhat conservative approach by focusing on $\wl\wl\to\wl\wl$
scattering, which is the most experimentally observable of the
channels that will display unitarity violation at large energies.

In discussing unitarity issues for $\wlwl$, we should note
that it is not necessary to consider the effects of the scalar
field(s) that are responsible for stabilizing the inter-brane
separation at the classical level. While these fields too will have
scalar excitations, the fields are normally chosen to be singlets
under the SM gauge groups (sample models include those of
Refs.~\cite{Goldberger:1999wh,Grzadkowski:2003fx}),
and will thus have no direct couplings to the $\wl\wl$ channel.  Their
effects through mixing with the Higgs and radion can be neglected.

As a final remark, we note that it would be interesting to analyze
unitarity constraints from KK graviton exchanges in other theories
with extra dimensions, such as the many models with flat extra
dimensions. We expect that the inclusion of the KK graviton modes
would significantly modify the constraints from unitarity that are
already known to arise from other types of KK excitations. For
example, in universal extra dimension models it is known that the KK
gauge boson excitations can cause unitarity violation if too many are
included~\cite{Chivukula:2003kq}.  Inclusion of the KK graviton
excitations could modify the situation.

%
\vskip .05in
\centerline{\bf Acknowledgments}
\smallskip

The authors thank Janusz Rosiek for his interest at the beginning of
this project.  B.G. thanks the CERN Theory Group for warm hospitality
during the period part of this work was performed.  This work is
supported in part by the Ministry of Science and Higher Education
(Poland) in years 2004-6 and 2006-8 as research projects 1~P03B~078~26
and N202~176~31/3844, respectively, by EU Marie Curie Research
Training Network HEPTOOLS, under contract MRTN-CT-2006-035505, by the
U.S. Department of Energy grant No. DE-FG03-91ER40674, and by NSF
International Collaboration Grant No. 0218130.  B.G. acknowledges the
support of the European Community under MTKD-CT-2005-029466 Project.
JFG thanks the Aspen Center for Physics where a portion of this work
was performed. JFG also thanks H.-C. Cheng, J. Lykken, B. McElrath and
J. Terning for helpful conversations.

\bigskip

\centerline{\bf APPENDIX: Rough derivations of some key results.}
\smallskip

We first give a rough derivation of why it is that $\lbar$ tracks
$\lnda$ and of the approximate numerical relation between them.  We
first recall that the spacing between KK states is, from
Eq.~(\ref{m1form}), roughly $\Delta m=\pi (\mompl){\lwh\over \sqrt
  2}$. From Eq.~(\ref{wwz}), at $\rts=\lbar$ we approximate the
contribution to $a_0$ of each KK state as $a_0^G\sim {12\over 384
  \pi\lwh^2}\lbar^2$, where we used $11+12 \log(m_G^2/\lbar^2)\sim
-12$ (which we have checked is fairly accurate as an average for all
KK states with mass below $\lnda$, nearly independent of $\mompl$) and
neglected all other terms. The number of contributing KK states is
roughly given by $N_G\sim {\lbar\over \Delta m}$. At the unitarity
limit we have $a_0={1\over 2}\sim a_0^G N_G$. Using this equation and
inputing the relation (\ref{nda}) between $\lwh$ and $\lnda$ yields
$\lbar={\lnda\over \pi^{1/3}}\sim 0.7 \lnda$, a result that is
remarkably close to that obtained via the complete calculation.

Second, we wish to estimate the value, $\mhmax\sim 1.43\tev$, of the
largest Higgs mass allowed, and why it is that this maximum is more or
less independent of $\mompl$.  We first note that in the extremal
situation being considered it turns out that unitarity is violated at
$\re a_0=-1/2$ at a $\rts$ value quite close to $\lbar$. One can
check this in the sample case presented in the right-hand plot of
Fig.~\ref{a012rts2}, where the $\mompl=0.0018$ curve touches $\re
a_0=-1/2$ at $\rts=3\tev$ as compared to $\lbar\sim
\lnda/\pi^{1/3}\sim 4.8\tev/\pi^{1/3}\sim 3.3\tev$.
Thus, for our estimate we consider $\rts=\lbar$ and, once again,
we use $11+12 \log(m_G^2/\lbar^2)\sim -12$ on average and sum over
$N_G\sim {\lbar\over \Delta m}$ KK states to obtain the net KK
contribution of $a_0^{KK}=+1/2$ (independent of $\mompl$), as employed
in deriving $\lbar$ above. We now add in the asymptotic contribution
of the Higgs boson for $\lbar\gg \mh$ as given in Eq.~(\ref{wwz}) of
$a_0^{h}=-{1\over 8\pi} {\mh^2\over v^2}$.  Setting
$a_0^{KK}+a_0^{h}=-1/2$ gives $\mh\sim 1.23\tev$, more or less
independent of $\mompl$.

The above derivation makes it clear that the $\lbar$ for obtaining
$\mhmax$ in the extremal situation is $\lbar\sim 3.3\tev$,
more or less independent of $\mompl$. Now, given that $\lbar\sim
\lnda/\pi^{1/3}$ and using Eq.~(\ref{nda}) with $\lwh=\lphi/\sqrt 3$,
one obtains $\mompl\sim (1.2\tev/\lphi)^3$ yielding values of
$\mompl\sim 1.38\times 10^{-2}$, $1.69\times 10^{-3}$, $2.16\times 10^{-4}$ and
$2.70\times 10^{-5}$ at $\lphi=5\tev$, $10\tev$, $20\tev$ and
$40\tev$, respectively.  These compare well with the values in the top
box of Table~\ref{primary}.


\begin{thebibliography}{99}
%
\vspace*{-.2in}

\bibitem{Randall:1999ee}
L.~Randall and R.~Sundrum,
Phys.\ Rev.\ Lett.\  {\bf 83} (1999) 3370
[arXiv:hep-ph/9905221];
%
Phys.\ Rev.\ Lett.\  {\bf 83} (1999) 4690
[arXiv:hep-th/9906064].
%
\bibitem{Dominici:2002jv}
D.~Dominici, B.~Grzadkowski, J.~F.~Gunion and M.~Toharia,
Nucl.\ Phys.\ B {\bf 671}, 243 (2003)
[arXiv:hep-ph/0206192];
%
Acta Phys.\ Polon.\ B {\bf 33}, 2507 (2002)
[arXiv:hep-ph/0206197].
%
\bibitem{Davoudiasl:1999jd}
  H.~Davoudiasl, J.~L.~Hewett and T.~G.~Rizzo,
  Phys.\ Rev.\ Lett.\  {\bf 84}, 2080 (2000)
  [arXiv:hep-ph/9909255].
%
\bibitem{Manohar:1983md}
  A.~Manohar and H.~Georgi,
  Nucl.\ Phys.\  B {\bf 234}, 189 (1984).
  Z.~Chacko, M.~A.~Luty and E.~Ponton,
  JHEP {\bf 0007}, 036 (2000)
  [arXiv:hep-ph/9909248].
%
%
\bibitem{Han:2004wt}
T.~Han and S.~Willenbrock,
Phys.\ Lett.\ B {\bf 616}, 215 (2005)
[arXiv:hep-ph/0404182].
%

%
\bibitem{Giudice:2000av}
G.~F.~Giudice, R.~Rattazzi and J.~D.~Wells,
Nucl.\ Phys.\ B {\bf 595}, 250 (2001)
[arXiv:hep-ph/0002178].
%
\bibitem{Grzadkowski:2005tx}
 Presented at the
XXIX International Conference of Theoretical Physics 
"Matter To The Deepest:
Recent Developments In Physics of Fundamental Interactions",
Ustron, 8-14 September 2005, Poland.,  B.~Grzadkowski and J.~Gunion,
  Acta Phys.\ Polon.\ B {\bf 36}, 3513 (2005).

\bibitem{Goldberger:1999wh}
  W.~D.~Goldberger and M.~B.~Wise,
  Phys.\ Rev.\ D {\bf 60}, 107505 (1999)
  [arXiv:hep-ph/9907218];
  Phys.\ Rev.\ Lett.\  {\bf 83}, 4922 (1999)
  [arXiv:hep-ph/9907447].

\bibitem{Grzadkowski:2003fx}
  B.~Grzadkowski and J.~F.~Gunion,
  Phys.\ Rev.\ D {\bf 68}, 055002 (2003)
  [arXiv:hep-ph/0304241].
%
\bibitem{Han:2001xs}
T.~Han, G.~D.~Kribs and B.~McElrath,
Phys.\ Rev.\ D {\bf 64}, 076003 (2001)
[arXiv:hep-ph/0104074].
%
\bibitem{Choudhury:2001ke}
D.~Choudhury, S.~R.~Choudhury, A.~Gupta and N.~Mahajan,
J.\ Phys.\ G {\bf 28}, 1191 (2002)
[arXiv:hep-ph/0104143];
%
U.~Mahanta,
arXiv:hep-ph/0004128.

%
\bibitem{Lee:1977eg}
B.~W.~Lee, C.~Quigg and H.~B.~Thacker,
Phys.\ Rev.\ D {\bf 16}, 1519 (1977).

\bibitem{magass} 
Carsten Magass, presentation on behalf of the D0 collaboration at DPF-2006, Honolulu, Hawaii.

\bibitem{Davoudiasl:2000wi}
  H.~Davoudiasl, J.~L.~Hewett and T.~G.~Rizzo,
  Phys.\ Rev.\ D {\bf 63}, 075004 (2001)
  [arXiv:hep-ph/0006041].
%
\bibitem{Csaki:2000zn}
  C.~Csaki, M.~L.~Graesser and G.~D.~Kribs,
  Phys.\ Rev.\  D {\bf 63}, 065002 (2001)
  [arXiv:hep-th/0008151].


\bibitem{Giudice:1998ck}
  G.~F.~Giudice, R.~Rattazzi and J.~D.~Wells,
  Nucl.\ Phys.\  B {\bf 544}, 3 (1999)
  [arXiv:hep-ph/9811291].
  G.~F.~Giudice and A.~Strumia,
  Nucl.\ Phys.\  B {\bf 663}, 377 (2003)
  [arXiv:hep-ph/0301232].
  G.~F.~Giudice, T.~Plehn and A.~Strumia,
  Nucl.\ Phys.\  B {\bf 706}, 455 (2005)
  [arXiv:hep-ph/0408320].

\bibitem{Dicus:2004rt}
  D.~Dicus and S.~Willenbrock,
  Phys.\ Lett.\ B {\bf 609}, 372 (2005)
  [arXiv:hep-ph/0409316].
%

\bibitem{Allanach:2002gn}
  B.~C.~Allanach, K.~Odagiri, M.~J.~Palmer, M.~A.~Parker, A.~Sabetfakhri and B.~R.~Webber,
  JHEP {\bf 0212}, 039 (2002)
  [arXiv:hep-ph/0211205].


\bibitem{Chivukula:2003kq}
  R.~S.~Chivukula, D.~A.~Dicus, H.~J.~He and S.~Nandi,
  Phys.\ Lett.\ B {\bf 562}, 109 (2003)
  [arXiv:hep-ph/0302263].

\end{thebibliography}
\end{document}